\documentclass[aip,pop,preprint]{revtex4-2}

\draft 

\usepackage{siunitx}
\usepackage{cancel}
\usepackage{amsmath}
\usepackage{amssymb}
\usepackage{mathtools}
\usepackage{amsthm}
\usepackage{amsfonts}
\usepackage[table, dvipsnames]{xcolor}
\usepackage{bm}
\usepackage{parskip}
\usepackage[margin=1in]{geometry}
\usepackage{graphicx}
\usepackage{listings}
\usepackage{hyperref}
\usepackage{multirow}
\usepackage{wrapfig}
\usepackage{subcaption}

\newcommand{\dx}{\,\mathrm{d} \bm{x}}
\newcommand{\dv}{\,\mathrm{d} \bm{v}}
\newcommand{\dtt}{\frac{\mathrm{d}}{\mathrm{d} t}}
\newcommand{\Alfven}{Alfv\'{e}n}
\newcommand{\vl}{v_\parallel}
\newcommand{\vp}{\bm{v}_\perp}
\newcommand{\np}{n_\perp}
\newcommand{\up}{\bm{u}_\perp}
\newcommand{\ups}{\bm{u}_\perp^*}
\newcommand{\Tp}{T_\perp}
\newcommand{\pp}{p_\perp}
\newcommand{\ep}{e_\perp}
\newcommand{\qp}{\bm{q}_\perp}
\newcommand{\Dp}{\nabla_\perp}
\newcommand{\Dl}{\mathcal{D}_\parallel}
\newcommand{\dvl}{\,\mathrm{d} \vl}
\newcommand{\dvp}{\,\mathrm{d} \vp}
\newcommand{\dldt}{\frac{\mathrm{d}^\parallel}{\mathrm{d} t}}
\newcommand{\dldtz}{\frac{\mathrm{d}^\parallel}{\mathrm{d} t^0}}
\renewcommand{\sp}{{s^\prime}}
\newcommand{\ssp}{{s s^\prime}}

\newcommand{\oct}{\omega_c \tau}
\newcommand{\opt}{\omega_p \tau}
\newcommand{\npt}{\nu_p \tau}

\newcommand{\npe}{n_{\perp e}}
\newcommand{\upe}{\bm{u}_{\perp e}}
\newcommand{\ppe}{p_{\perp e}}
\newcommand{\qpe}{\bm{q}_{\perp e}}
\newcommand{\epe}{e_{\perp e}}

\definecolor{tablerowgray}{gray}{0.9}

\begin{document}


\title{Asymptotic perpendicular transport in low-beta collisionless plasma} 



\author{Jack Coughlin}
\email{jack@johnbcoughlin.com}
\author{Jingwei Hu}
\email{hujw@uw.edu}
\affiliation{Department of Applied Mathematics, University of Washington, Seattle, WA 98195, USA.}
\author{Uri Shumlak}
\email{shumlak@uw.edu}
\affiliation{Aerospace and Energetics Research Program, University of Washington, Seattle, WA 98195, USA.}


\date{August 12, 2024}

\begin{abstract}
Kinetic physics, including finite Larmor radius (FLR) effects, are known to affect the physics of magnetized plasma phenomena such as the Kelvin-Helmholtz and Rayleigh-Taylor instabilities. Accurately incorporating FLR effects into fluid simulations requires moment closures for the heat flux and stress tensor, including the gyroviscous stress in collisionless magnetized plasmas. However, the most commonly used gyroviscous stress tensor closure (Braginskii Rev. Plasma Phys., 1965) is based on a strongly collisional assumption for the asymptotic expansion of the kinetic equa- tion in the so-called fast-dynamics ordering. This collisional assumption becomes less valid for some high-temperature plasmas. To explore perpendicular transport in collisionless and weakly collisional plasmas, an asymptotic analysis of the weakly collisional Vlasov equation in the slow-dynamics or drift ordering is performed in a new “semi-fluid” formalism, which integrates in $\bm{v}_\perp$ to obtain a five-moment system. The associated heat flux and stress tensor closures are determined via a Hilbert expansion of the kinetic equation. A numerically affordable approximation to the stress tensor is proposed which adjusts the Braginskii closure to account for temperature gradient-driven stress. Continuum kinetic simulations of a family of sheared-flow configurations with variable magnetization and temperature gradients are performed to validate the drift ordering semi-fluid expansion. The expected convergence with magnetization is observed, and residuals are examined and discussed in terms of their relationship to higher-order terms in the expansion. The adjusted Braginskii closure is found to accurately correct for the error committed by the Braginskii gyroviscous stress tensor closure in the presence of temperature gradients.
\end{abstract}


\maketitle 

\section{Introduction}

The transport of plasma particles, momentum and energy across confining magnetic field lines is
of interest in many different plasma applications.
Low-beta plasmas, in which the plasma pressure is much lower than the magnetic field pressure, 
are ubiquitous in magnetic confinement fusion concepts including tokamaks and stellarators.
In these plasmas, cross-field transport is known to be affected by gradients of density and temperature
perpendicular to the magnetic field.
This perpendicular transport is mediated by non-ideal effects such as collisions
as well as by unstable interchange modes.

Ideal transport, which is captured by models such as ideal MHD and
the ideal five-moment two-fluid (5M2F) model, is distinguished from non-ideal transport.
Non-ideal transport effects include viscosity, heat diffusion, and resistivity.
Collisions can be a major contributor to perpendicular non-ideal transport in magnetized plasmas.
However, in many regimes of interest the collision-driven perpendicular transport is slow enough
as to be insignificant compared to collisionless physics.
In the absence of collisions, finite-Larmor radius (FLR) effects are known to contribute to viscosity 
through the gyroviscous stress \cite{braginskiiTransportProcessesPlasma1965, ramosGeneralExpressionGyroviscous2005}.
The effects of gyroviscous momentum transport on fluid instabilities have been extensively investigated
through theory and simulation.
Examples of studied instabilities include the magnetized Rayleigh-Taylor 
\cite{hubaFiniteLarmorRadius1996, hubaNonlocalTheoryRayleigh1989, hubaTheorySimulationRayleighTaylor1987}
and Kelvin-Helmholtz instabilities
\cite{umedaEvaluatingGyroviscosityKelvinHelmholtz2016, vogmanTwofluidKineticTransport2020, 
vogmanHighfidelityKineticModeling2021}.

From this discussion, it is clear that accurately capturing non-ideal transport physics
is important for accurate modeling of plasma evolution.
The most physically accurate model of collisionless plasmas is provided by kinetic theory.
However, the theoretical and computational difficulties posed by kinetic theory motivates
research into reduced models based on fluid equations which can also capture kinetic effects such as 
FLR effects.
In the context of collisionless plasmas, the fluid equations require closures for the
flux moments of heat and momentum, which are respectively the heat flux vector and the stress tensor.

The most widely cited such closure is the Braginskii 5M2F model \cite{braginskiiTransportProcessesPlasma1965}.
Braginskii developed diffusive closures for the ion and electron heat flux, stress tensor,
and resistivity based on a Chapman-Enskog type expansion around a Maxwellian equilibrium.
The Maxwellian equilibrium, and subsequent development of the asymptotic expansion, is
based on the assumption of strong collisions described by a 
Landau-Fokker-Planck collision operator \cite{landauTRANSPORTEQUATIONCASE1965}.
Formally, the Braginskii closure is valid in the strongly collisional and magnetized limit,
$\nu_p \tau, \omega_c \tau \rightarrow \infty$, where $\tau$ is the characteristic time scale,
$\nu_p$ the proton collision frequency and $\omega_c$ the proton cyclotron frequency.
Within this regime, one can consider the relative magnetization $\omega_c / \nu_p$.
The strongly magnetized limit $\omega_c / \nu_p \rightarrow \infty$ can be taken and gives 
meaningful closures for the transport terms which are independent of collision frequency, 
namely the diamagnetic heat flux $\bm{q}_\wedge$ and the gyroviscous stress tensor $\Pi_\wedge$.

The Braginskii closure correctly predicts the physical phenomenon of gyroviscosity and gives
a tractable closure for numerical implementation.
However, the assumption of strong collisions is unsatisfactory for plasmas in the collisionless regime.
For this reason, collisionless and weakly collisional fluid closures for magnetized plasmas have also been developed 
\cite{chewBoltzmannEquationOnefluid1956, ramosFluidFormalismCollisionless2005, ramosFluidTheoryMagnetized2007, macmahonFiniteGyroRadiusCorrections1965, zhengPerpendicularMagnetofluidTheory2020, ramosGeneralExpressionGyroviscous2005, simakovMomentumTransportArbitrary2007}.

In this work we develop a new collisionless fluid closure for perpendicular transport
of low-beta magnetized plasmas in the drift ordering. 
The closure theory is distinguished from previous work 
\cite{ramosFluidFormalismCollisionless2005, ramosFluidTheoryMagnetized2007, 
macmahonFiniteGyroRadiusCorrections1965}
by first reducing the Vlasov equation to an intermediate set of ``semi-fluid'' equations
for the perpendicular velocity moments only, leaving the parallel velocity dependence kinetic.
In this respect the approach is similar to Ref. \onlinecite{zhengPerpendicularMagnetofluidTheory2020},
    although that reference uses the same ``fast-dynamics'' ordering as the Braginskii closure.
The closure developed here requires only the assumption of a strong magnetic field and that
the corresponding leading-order distribution function, which in the chosen ordering
necessarily has zero drift velocity, also have a Maxwellian distribution of perpendicular
particle energies.
Based on these assumptions, FLR effects are calculated to leading significant order
and found to include diamagnetic heat flux and gyroviscous stress terms.
Collisional terms are retained in an abstract form on the right-hand side to indicate how
the expansion can be extended to include collisional effects.
The resulting closure for heat flux is similar to Braginskii's diamagnetic heat flux closure,
while the gyroviscous stress tensor to leading order includes terms associated with the
temperature gradient.

To verify the transport theory numerically and to better understand the conditions of its
validity, we perform continuum kinetic simulations of the Vlasov equation.
Simulations are focused on a family of initial conditions which are both designed to 
elicit the transport phenomena predicted by theory and are of inherent physical interest:
cross-field sheared flow with density and temperature gradients.
The physics of sheared plasma flows has been studied extensively in numerical simulation
\cite{vogmanTwofluidKineticTransport2020, meierDevelopmentFivemomentTwofluid2021},
and is thought to be fundamental to the stabilization of magnetic confinement fusion
configurations such as the sheared-flow stabilized Z-pinch 
\cite{shumlakZpinchFusion2020, shumlakIncreasingPlasmaParameters2017}
and the H-mode confinement regime in tokamaks \cite{keilhackerHmodeConfinementTokamaks1987}.
The kinetic heat flux and stress tensor are compared to the predictions of our closure
and of Braginskii's closure for the diamagnetic heat flux and gyroviscous stress tensor
over a range of magnetizations and temperature gradients.
    Braginskii's closure is found to neglect a contribution to the stress tensor from temperature
    gradients, of significance in the drift ordering regime, which our closure accurately incorporates.
    This contribution is approximated by a numerically affordable adjustment to the Braginskii
    closure. 
    The resulting adjusted Braginskii closure more accurately captures the leading-order 
    stress tensor physics in the presence of temperature gradients.
    Additionally, contour plots of the residuals of the transport closures are analyzed,
    and found to be consistent with the presence of second- and third-order effects as expected
    from asymptotic theory.

This paper is organized as follows.
Section \ref{sec:fluid_models} covers background on the Vlasov
equation, the classical derivation of the 5M2F model,
and the collisionless limit of the Braginskii closure.
Section \ref{sec:scaling_assumptions} gives a rigorous presentation
of the asymptotic scaling assumptions that we base our subsequent derivation on.
Section \ref{sec:semi_fluid_closure} derives the semi-fluid equations and
our asymptotic transport theory.
Section \ref{sec:kinetic_modeling} contains a description of the continuum kinetic
code we use to validate the results of Section \ref{sec:semi_fluid_closure} and of
the class of initial conditions that we consider here. 
Section \ref{sec:numerical_results} describes the results of our computational experiments.
Finally, Section \ref{sec:conclusion} closes with a discussion of the newly derived
transport theory as it relates to existing theories.

\section{Kinetic and fluid models of plasma}
\label{sec:fluid_models}

Magnetically confined plasmas are described by a hierarchy of model equations.
The most fundamental is the Vlasov equation, which for a species $s$ is written
\begin{align}
    \label{eqn:vlasov}
    \partial_t f_s + \bm{v} \cdot \nabla_{\bm{x}} f_s + \frac{q_s}{m_s} \left[ \bm{E} + \bm{v} \times \bm{B} \right] \cdot \nabla_{\bm{v}} f_s = C(f_s),
\end{align}
where $q_s$ is the species charge, $m_s$ its mass, and $\bm{E}$ and $\bm{B}$ the electric and magnetic
fields, respectively.
The unknown $f_s(\bm{x}, \bm{v}, t)$ is the particle distribution function which we take to have units of
$\SI{}{m^{-6} s^{3}}$ so that its zeroth velocity moment is $n(\bm{x}, t)$, the number density.
The collision term $C(f_s)$ on the right-hand side captures all collisions, and may be modeled
using any of a wide variety of collision operators. 
The most physically accurate is generally considered to be the Landau-Fokker-Planck operator 
for Coulomb collisions \cite{landauTRANSPORTEQUATIONCASE1965}.
In what follows, we will use the un-subscripted notation for the spatial gradient: $\nabla = \nabla_{\bm{x}}$.

The Vlasov equation is coupled to Maxwell's equations for the electromagnetic fields:
\begin{align}
    \label{eqn:gauss}
    \nabla \cdot \bm{E} &= \frac{\rho_c}{\epsilon_0}  \\
    \label{eqn:divB}
    \nabla \cdot \bm{B} &= 0 \\
    \label{eqn:faraday}
    \nabla \times \bm{E} &= -\frac{\partial \bm{B}}{\partial t} \\
    \label{eqn:ampere}
    \nabla \times \bm{B} &= \mu_0 \left( \bm{j} + \epsilon_0 \frac{\partial \bm{E}}{\partial t} \right),
\end{align}
where $\epsilon_0$ and $\mu_0$ are the permittivity and permeability of vacuum, respectively, and
\begin{align*}
\rho_c = \sum_s q_s \int f_s \dv, \quad \bm{j} = \sum_s q_s \int \bm{v} f_s \dv
\end{align*}
are the charge and current density of the plasma.

Low-beta plasmas with a constant applied magnetic field can be well-approximated by 
the electrostatic approximation \cite{kadomtsevTurbulentDiffusionRarefied1964}, 
which takes $\partial_t \bm{B} = 0$ and solves
for $\bm{E}$ via the Poisson equation
\begin{align*}
\bm{E} = -\nabla \phi, \quad \nabla^2 \phi = -\frac{\rho_c}{\epsilon_0}.
\end{align*}

The 5M2F fluid model may be derived by taking velocity moments of \eqref{eqn:vlasov}.
Define a moment-taking operator $\left\langle \psi(\bm{v}), \cdot \right\rangle_v$ by
\begin{align*}
    \left\langle \psi(\bm{v}), \cdot \right\rangle_v = \int_{\mathbb{R}^3} \psi(\bm{v}) \cdot \dv.
\end{align*}
The ideal 5M2F model is derived by taking the moments $\left\langle 1, \cdot \right\rangle_v, m_s \left\langle \bm{v}, \cdot \right\rangle_v, \frac{m_s}{2}\left\langle |\bm{v}|^2, \cdot \right\rangle_v$ of \eqref{eqn:vlasov}
with vanishing right-hand side:
\begin{align}
    \label{eqn:5M-continuity}
&\partial_t n_s + \nabla \cdot (n_s \bm{u}_s) = 0 \\
    \label{eqn:5M-momentum}
m_s&\partial_t (n_s \bm{u}_s) + \nabla \cdot (m_s n_s \bm{u}_s \otimes \bm{u}_s + \mathbb{P}_s) = n_s q_s (\bm{E} + \bm{u}_s \times \bm{B}) \\
    \label{eqn:5M-energy}
   &\partial_t e_s + \nabla \cdot \left((e_s \mathbb{I} + \mathbb{P}_s) \cdot \bm{u}_s + \bm{q}_s\right) = n_s q_s \bm{E} \cdot \bm{u}_s
\end{align}
The number density $n_s$, velocity $\bm{u}_s$, species energy $e_s$, pressure tensor $\mathbb{P}_s$,
and heat flux $\bm{q}_s$ are defined by the following moments of $f_s$:
\begin{align}
\label{eqn:def-n-nu-e}
&n_s = \left\langle 1, f_s \right\rangle_v, \quad n_s \bm{u}_s = \left\langle \bm{v}, f_s \right\rangle_v, \quad e_s = \frac{m_s}{2}\left\langle |\bm{v}|^2, f_s \right\rangle_v \\
\label{eqn:def-Pbb-q}
&\mathbb{P}_s = m_s \left\langle (\bm{v} - \bm{u}_s) \otimes (\bm{v} - \bm{u}_s), f_s \right\rangle_v \quad \bm{q}_s = \frac{m_s}{2} \left\langle (\bm{v} - \bm{u}_s) |\bm{v} - \bm{u}_s|^2, f \right\rangle_v.
\end{align}
The pressure tensor can be split into a diagonal component and a trace-free component:
\begin{align*}
\mathbb{P}_s = p_s \mathbb{I} + \Pi_s,
\end{align*}
where $\mathbb{I}$ is an identity tensor,
\begin{align}
    \label{eqn:pressure_equation_of_state}
p_s = (\gamma - 1) \left(e_s - \frac{m_s n_s |\bm{u}_s|^2}{2}\right)
\end{align}
is the familiar scalar pressure, with $\gamma = 5/3$ the ratio of specific heats.
The scalar temperature is defined via the scalar pressure as $T_s = p_s / n_s$.
The trace-free component $\Pi_s$ is known as the stress tensor.
Equation \eqref{eqn:pressure_equation_of_state} gives a closed-form expression for $p_s$
in terms of the conserved quantities evolved by \eqref{eqn:5M-continuity}, \eqref{eqn:5M-momentum}
and \eqref{eqn:5M-energy}.
The undetermined moments appearing in the 5M2F model are therefore $\bm{q}_s$ and $\Pi_s$.

\subsection{Collisionless limit of Braginskii ion closures}

\subsubsection*{Heat flux}
The Braginskii \cite{braginskiiTransportProcessesPlasma1965} expression for the ion heat flux is
\begin{align}
\bm{q}_i = -\kappa^i_\parallel \nabla_\parallel T_i - \kappa^i_\perp \nabla_\perp T_i + \kappa^i_\wedge \hat{\bm{b}} \times \nabla T_i.
\end{align}
The unit vector $\hat{\bm{b}}$ is defined by $\hat{\bm{b}} = \bm{B} / |B|$.
The parallel and perpendicular gradient operators are defined relative to $\hat{\bm{b}}$, i.e. $\nabla_\parallel = \hat{\bm{b}} (\hat{\bm{b}} \cdot \nabla)$ and $\nabla = \nabla_\perp + \nabla_\parallel$.
The ion heat conductivities are
\begin{align*}
    \kappa^i_\parallel &= 3.906 \frac{p_i}{m_i \nu_i}, \\
    \kappa^i_\perp &= \frac{2x^2 + 2.645}{\Delta} \frac{p_i}{m_i \nu_i}, \\
    \kappa^i_\wedge &= \frac{2.5 x^3 + 4.65x}{\Delta} \frac{p_i}{m_i \nu_i},
\end{align*}
where $\nu_i$ is the ion collision frequency and $x = \omega_{ci} / \nu_i$ is the magnetization parameter,
and $\Delta = x^4 + 2.7 x^2 + 0.677$.
The collisionless limit is represented by $\nu_i \rightarrow 0$.
In this limit the perpendicular heat conductivity vanishes, while the diamagnetic heat
conductivity becomes
\begin{align*}
    \kappa^i_\wedge \rightarrow \frac{5 p_i}{2 m_i \omega_{ci}}.
\end{align*}
The Braginskii estimate of the heat flux in the collisionless limit is therefore
\begin{align*}
    \bm{q}_i^{Brag} = \frac{5 p_i}{2 m_i \omega_{ci}} \hat{\bm{b}} \times \nabla T_i.
\end{align*}
The parallel heat conductivity becomes infinite as $\nu_i \rightarrow 0$, which is a clearly
unphysical result, but this paper
is primarily concerned with plasmas which can be considered symmetric in the parallel
direction, so we do not treat parallel heat conduction.

\subsubsection*{Stress tensor}
The expression for the ion stress tensor is
\begin{align}
\Pi_i = -\eta_0 \mathbb{W}_0 - \eta_1 \mathbb{W}_1 - \eta_2 \mathbb{W}_2 + \eta_3 \mathbb{W}_3 + \eta_4 \mathbb{W}_4.
\end{align}
The definitions of the tensors $\mathbb{W}_{0\dots 4}$ can be found in 
\cite{braginskiiTransportProcessesPlasma1965} (4.42).
The viscosity coefficients are given by
\begin{align*}
\eta_0 = \frac{0.96 p_i}{\nu_i}, \quad \eta_2 = \frac{\frac{6}{5}x^2 + 2.23}{\Delta} \frac{p_i}{\nu_i},
\quad \eta_4 = \frac{x^3 + 2.38 x}{\Delta}\frac{p_i}{\nu_i},
\end{align*}
and
\begin{align*}
\eta_1 = \eta_2(2x), \quad \eta_3 = \eta_4(2x).
\end{align*}
Taking the limit $\nu_i \rightarrow 0$, both $\eta_1$ and $\eta_2$ vanish, which is consistent
with the physical picture that perpendicular viscous stress is driven by collisional processes.
On the other hand, $\eta_0$, which is associated with stress due to elongation of the distribution
function in the parallel direction, goes to infinity, indicating that the Braginskii closure gives
an unphysical solution in the strongly magnetized (or weakly collisional) limit.
As with the parallel heat flux, in this paper we are not concerned with viscous stress due 
to parallel elongation of the distribution.
The transport coefficients $\eta_3$ and $\eta_4$ have finite limits, which are
\begin{align*}
    \eta_3 = \frac{p_i}{2\omega_{ci}}, \quad \eta_4 = \frac{p_i}{\omega_{ci}}.
\end{align*}
In the case of a symmetric plasma in the parallel direction the term with coefficient $\eta_4$
vanishes since it is associated with parallel components of the stress tensor.
It only remains to consider the term with coefficient $\eta_3$.
For notational simplicity, consider the case where the magnetic field is oriented in the $z$
direction.
In the strongly magnetized limit, the perpendicular ($xy$) components of the stress tensor are
\begin{align}
    \label{eqn:Pi_gyro_Brag}
    \Pi_\perp^{Brag} = \frac{p_i}{2 \omega_{ci}} \mathbb{W}_3 = \frac{p_i}{2 \omega_{ci}} \begin{pmatrix}
        -\mathbb{W}_{xy} & \frac{1}{2} (\mathbb{W}_{xx} - \mathbb{W}_{yy}) \\
        \frac{1}{2} (\mathbb{W}_{xx} - \mathbb{W}_{yy}) & \mathbb{W}_{xy},
    \end{pmatrix}
\end{align}
where
\begin{align*}
\mathbb{W} = \nabla \bm{u}_i + (\nabla \bm{u}_i)^T - \frac{2 \mathbb{I}}{3} \nabla \cdot \bm{u}_i
\end{align*}
is the shear stress tensor.
The perpendicular stress tensor given by \eqref{eqn:Pi_gyro_Brag} is known as the gyroviscous stress tensor.
It is associated with transport of $x$-momentum in the $y$ direction and vice versa.
A notable property of the gyroviscous stress tensor is that it does not contribute to dissipative viscous heating, since
\begin{align*}
\mathbb{W}_3 : \nabla \bm{u}_i = 0,
\end{align*}
and thus the corresponding term in the non-conservative temperature equation vanishes:
\begin{align*}
\dtt T_i + \frac{\gamma-1}{n_i} \left( \mathbb{P}_i : \nabla \bm{u}_i + \nabla \cdot \bm{q}_i \right) = 0,
\end{align*}
where $\dtt = \partial_t + \bm{u}_i \cdot \nabla$ denotes the material derivative.

\section{Scaling assumptions}
\label{sec:scaling_assumptions}

In this section we make precise our normalization and scaling assumptions.
Our normalization is based on the flexible plasma normalization described in Ref.
\onlinecite{millerMultispecies13momentModel2016}.
Beginning with the dimensional Vlasov equation for species $s$, \eqref{eqn:vlasov},
the species charge and mass are normalized by the proton charge $e$ and mass $m_p$:
\begin{align*}
m_s = A_s m_p, \quad q_s = Z_s e.
\end{align*}
The reference proton plasma frequency is given by
\begin{align*}
\omega_p^2 = \frac{e^2 n_0}{m_p \epsilon_0} 
\end{align*}
where $n_0$ is a reference number density.
The plasma frequency eliminates $\epsilon_0$, while $\mu_0$ is eliminated by the introduction of the reference
\Alfven\ velocity
\begin{align*}
    v_A^2 = \frac{B_0^2}{m_p n_0 \mu_0}.
\end{align*}
The reference velocity is set to $v_0 = v_A$.
We introduce characteristic length and time scales via
\begin{align*}
    \bm{x} = L \overline{\bm{x}}, \quad t = \tau \overline{t},
\end{align*}
where $\tau$ is a reference timescale and $L = v_0 \tau$.
Finally, reference phase space densities and collision operators are introduced via
\begin{align*}
f_s = f_0 \overline{f}_s, \quad C(f_s) = \nu_p f_0 \overline{C}(\overline{f}_s).
\end{align*}

The reference quantities are used to nondimensionalize \eqref{eqn:vlasov} by substituting
expressions such as $\bm{v} = v_0 \overline{\bm{v}}$, where the notational convention is that
overlined quantities are of order unity.
Doing so gives
\begin{align*}
    \frac{f_0}{\tau}\partial_{\overline{t}} \overline{f}_s + \frac{f_0 v_0}{L}\overline{\bm{v}} \cdot \overline{\nabla} \overline{f}_s + \frac{Z_s e}{A_s m_p} \left[ \frac{m_p v_0 \omega_p}{e} \overline{\bm{E}} + v_0 B_0 (\overline{\bm{v}} \times \overline{\bm{B}}) \right] \cdot \frac{f_0}{v_0} \nabla_{\overline{\bm{v}}} \overline{f}_s = \nu_p f_0 \overline{C}(\overline{f}_s).
\end{align*}
Multiplying through by $\tau / f_0$ completes the nondimensionalization:
\begin{align}
    \label{eqn:vlasov_normalized-1}
    \partial_{\overline{t}}\overline{f}_s + \overline{\bm{v}} \cdot \nabla \overline{f}_s + \frac{Z_s}{A_s} \left[ \omega_p \tau \overline{\bm{E}} + \omega_c \tau (\overline{\bm{v}}\times \overline{\bm{B}}) \right] \cdot \nabla_{\overline{\bm{v}}} \overline{f}_s = \nu_p \tau \overline{C}(\overline{f}_s),
\end{align}
where $\omega_c = B_0 e / m_p$ is the reference proton cyclotron frequency.

The overlined quantites in \eqref{eqn:vlasov_normalized-1} are dimensionless, but not necessarily of
order unity.
To make our scaling assumption explicit, we define the small parameter $\epsilon = \tilde{\bm{B}} / \overline{\bm{B}}$
relating the dimensionless $\overline{\bm{B}}$ to the order-unity $\tilde{\bm{B}}$.
All other dimensionless quantities are assumed to be of order unity, so that e.g. $\tilde{\bm{v}} = \overline{\bm{v}}$.
Substituting for order-unity unknowns gives the Vlasov equation in the strongly magnetized scaling,
\begin{align}
    \label{eqn:vlasov_normalized}
    \partial_t f_s + \bm{v} \cdot \nabla f_s + \frac{Z_s}{A_s} \left[ \omega_p \tau \bm{E} + \epsilon^{-1} \omega_c \tau (\bm{v} \times \bm{B}) \right] \cdot \nabla_{\bm{v}} f_s = \nu_p \tau C(f_s).
\end{align}
Tildes are omitted in \eqref{eqn:vlasov_normalized} and in all subsequent expressions for clarity.

Equation \eqref{eqn:vlasov_normalized} is expressed in the flexible normalization form, which is characterized
by three dimensionless parameters $\omega_p \tau$, $\omega_c \tau$, and $\nu_p \tau$.
These characterize the strength of electrostatic forces, magnetic forces, and collisions, respectively.
Equation \eqref{eqn:vlasov_normalized} is additionally equipped with a formal small parameter $\epsilon$
around which we will perform asymptotic expansion in the following section.

Before proceeding, we first write some important plasma parameters in terms of $\epsilon$.
The reference plasma temperature is defined as $T_0 = \epsilon^2 m_p v_0^2$, and the reference pressure $p_0 = n_0 T_0$.
Thus the plasma beta scales as $\epsilon^2$:
\begin{align*}
\beta_0 = \frac{p_0}{B_0^2 / 2 \mu_0} = \frac{n_0 m_p \epsilon^2 v_A^2}{B_0^2 / 2 \mu_0} = 2\epsilon^2.
\end{align*}
The nondimensional proton Larmor radius is
\begin{align*}
    \frac{r_{Li}}{L} = \frac{\sqrt{T_0 / m_p}}{\omega_c L} = \frac{\epsilon}{\omega_c \tau}.
\end{align*}

\section{Semi-fluid model and collisionless magnetized closure}
\label{sec:semi_fluid_closure}

In this section we derive the set of semi-fluid equations and their leading-order transport 
closures from \eqref{eqn:vlasov_normalized}.
The derivation is based on the assumption of uniform $\bm{B}$ and straight field lines.
Thus, $\hat{\bm{b}}$ is a constant unit vector.
The phase space variables split into parallel and perpendicular components with respect to $\hat{\bm{b}}$:
\begin{align*}
    x_\parallel &= \bm{x} \cdot \hat{\bm{b}}, \quad \bm{x}_\perp = -(\bm{x} \times \hat{\bm{b}}) \times \hat{\bm{b}} \\
    v_\parallel &= \bm{v} \cdot \hat{\bm{b}}, \quad \bm{v}_\perp = -(\bm{v} \times \hat{\bm{b}}) \times \hat{\bm{b}}.
\end{align*}
The gradient operator also splits into parallel and perpendicular components:
\begin{align*}
\nabla_\parallel = \hat{\bm{b}} (\hat{\bm{b}} \cdot \nabla) = \hat{\bm{b}} \partial_{x_\parallel}, \quad \Dp = \nabla_{\bm{x}_\perp}
\end{align*}

The collisionless perpendicular transport theory derived here is based on a set of reduced equations,
which we call semi-fluid equations, which are obtained by taking moments
of \eqref{eqn:vlasov_normalized} in $\vp$.
This derivation results in semi-fluid analogues of the usual five-moment fluid equations 
\eqref{eqn:5M-continuity}-\eqref{eqn:5M-energy}.
The semi-fluid equations are PDEs posed over $\bm{x}, t,$ and unconventionally, $\vl$.
The semi-fluid moment hierarchy presents the typical moment-closure problem: by cutting off the hierarchy after
the energy equation, the perpendicular flux of (perpendicular) heat, and the full perpendicular pressure
tensor are undetermined.
The closure problem is addressed in the usual way by introducing a Hilbert expansion for $f_s$ centered 
around a gyrotropic Maxwellian.
Higher-order corrections are obtained by inverting the leading-order operator which in the case of
\eqref{eqn:vlasov_normalized} is the $\bm{v} \times \bm{B}$ force.

The asymptotic limit considered here is the drift ordering, which assumes that the perpendicular drift velocity
$\bm{u}_\perp$ satisfies $|\bm{u}_\perp| \ll v_{t}$ where $v_{t}$ is the thermal velocity.
As we will see this is a necessary consequence of the scaling assumptions made in Section \ref{sec:scaling_assumptions},
since the leading-order distribution $f^0$ must be gyrotropic.
The leading-order drift velocity $\bm{u}$ therefore appears as a moment of $f^1$,
and at the same order as the heat flux $\bm{q}$. 
This is an important difference from the Braginskii, or fast dynamics ordering, and as we will see 
it has consequences for the gyroviscous stress closure.
Another difference from the Braginskii ordering and asymptotic expansion is that the collision operator appears explicitly, i.e.
on the right-hand side, of each subsequent correction equation.
This makes it quite simple to accomodate different model collision operators such as the full Landau-Fokker-Planck
operator in the expansion.

\subsection{Semi-fluid equations}
The semi-fluid equations are a system of equations for a set of semi-fluid moments, which are obtained
by taking perpendicular velocity moments of $f_s$.
In these and most equations that follow we omit the species subscript $s$.
\begin{align}
    \np(\bm{x}, \vl, t) &= \int f \dvp,  \\
    \up(\bm{x}, \vl, t) &= \frac{1}{\np} \int \vp f \dvp, \\
    \ep(\bm{x}, \vl, t) &= \frac{A}{2} \int |\vp|^2 f \dvp, \\
    \mathbb{P}_\perp(\bm{x}, \vl, t) &= A \int (\vp - \up) \otimes (\vp - \up) f \dvp, \\
    \bm{q}_\perp(\bm{x}, \vl, t) &= \frac{A}{2} \int (\vp - \up) |\vp - \up|^2 f \dvp, \\
    \mathcal{N}(\bm{x}, \vl, t) &= \int C(f)\dvp, \\
    \mathcal{S}(\bm{x}, \vl, t) &= A \int \vp C(f) \dvp, \\
    \mathcal{Q}(\bm{x}, \vl, t) &= \frac{A}{2} \int |\vp|^2 C(f) \dvp.
\end{align}
The perpendicular scalar pressure and temperature are
given by $\pp = \frac{\text{Tr}(\mathbb{P}_\perp)}{2}$ and $\Tp = \pp / \np$.
By taking the zeroth moment of \eqref{eqn:vlasov_normalized} in $\vp$ we get the semi-fluid continuity equation:
\begin{align}
    \label{eqn:semifluid-continuity}
\dldt \np + \Dp \cdot (\np \up) = \npt \mathcal{N},
\end{align}
where the total derivative in the parallel direction is defined $\dldt = \partial_t + \mathcal{D}_\parallel$,
with $\mathcal{D}_\parallel$ the parallel Vlasov operator
\begin{align*}
    \Dl = \vl \partial_\parallel + \frac{Z}{A} E_\parallel \partial_{\vl}.
\end{align*}
Equation \eqref{eqn:semifluid-continuity} describes the evolution of the density of particles with a given
parallel velocity in space.
It resembles the fluid continuity equation \eqref{eqn:5M-continuity} in the perpendicular direction, but
in the parallel direction its dynamics are governed by a Vlasov operator.
It is important to note the presence of the source term $\mathcal{N}(\bm{x}, \vl, t)$ on the right-hand side, which is the 
zeroth $\vp$ moment of the collision term. It represents particles which are scattered to or away from a given
parallel velocity by collisions.
As such, it must satisfy an overall particle conservation property, which is
\begin{align*}
\int \mathcal{N} \dvl = 0.
\end{align*}

The perpendicular momentum equation is obtained by taking the first $\vp$ moment of \eqref{eqn:vlasov_normalized}, giving
\begin{align}
    \label{eqn:semifluid-momentum}
A \dldt(\np \up) + \Dp \cdot (A \np \up \otimes \up + \mathbb{P}_\perp) = \np Z (\omega_p \tau \bm{E} + \epsilon^{-1} \omega_c \tau \up \times \bm{B}) + \npt \mathcal{S}.
\end{align}
The flux term of equation \eqref{eqn:semifluid-momentum} contains the familiar full pressure tensor $\mathbb{P}_\perp$.
In the five-moment semi-fluid system considered here, the trace-free part of $\mathbb{P}_\perp$ requires a closure
relation, just as in the classical five-moment fluid system.
Equation \eqref{eqn:semifluid-momentum}
also contains a collisional momentum source term $\mathcal{S}(\bm{x}, \vl, t)$, 
which represents a source of perpendicular momentum at 
the given parallel velocity coordinate.
As such, it contains contributions from particles scattering into or away from $\vl$, as well as contributions
from cross-species exchange of perpendicular momentum at a given $\vl$.

The perpendicular energy equation is obtained by taking the moment of \eqref{eqn:vlasov_normalized} with respect to $A |\vp|^2/2$:
\begin{align}
    \label{eqn:semifluid-energy}
    \dldt \ep + \Dp \cdot ((\ep \mathbb{I} + \mathbb{P}_\perp) \cdot \up + \qp) = \np Z \opt \bm{E} \cdot \up + \npt \mathcal{Q}.
\end{align}
The flux term includes the second unclosed moment for the five-moment semi-fluid system, namely
$\qp$, the perpendicular heat flux.
Note that $\qp$ represents the perpendicular flux of thermal energy due to random perpendicular velocities,
but not random parallel velocities.
This is in contrast to the usual heat flux vector $\bm{q}$, whose perpendicular components
include the flux of thermal energy due to random particle velocities in all 3 dimensions.
Equation \eqref{eqn:semifluid-energy} also has a collisional source term on the right hand side, 
$\mathcal{Q}(\bm{x}, \vl, t)$,
which contains contributions from the energy of particles scattering into or away from $\vl$
as well as contributions from cross-species exchange of energy due to collisions.

\subsection{Hilbert expansions}
To calculate closures for $\mathbb{P}_\perp$ and $\bm{q}_\perp$, we introduce a Hilbert expansion for $f$
in terms of $\epsilon$:
\begin{align}
    \label{eqn:hilbert-expansion-f}
f = f^0 + \epsilon f^1 + \epsilon^2 f^2 + \cdots.
\end{align}
The macroscopic fluid variables can also be equipped with a Hilbert expansion.
For example,
\begin{align*}
\pp = \pp^0 + \epsilon \pp^1 + \epsilon^2 \pp^2 + \cdots.
\end{align*}
In order to leave the treatment of the collision terms until later, it is convenient
to supply a Hilbert expansion for the collisional term,
\begin{align}
    \label{eqn:hilbert-expansion-C}
C(f) = C^0(f) + \epsilon C^1(f) + \epsilon^2 C^2(f) + \cdots.
\end{align}
The details of how a given collision operator splits into an expansion such as \eqref{eqn:hilbert-expansion-C}
when acting on \eqref{eqn:hilbert-expansion-f} must be determined.
However, the specific form of the collision operator does not make a difference to the
derivation in the collisionless limit which is the focus of this paper.

Finally, medium and slow time scales are introduced by letting $t = t^0 + \epsilon^{-1} t^1$,
in terms of which the time derivative expands as
\begin{align*}
    \partial_t = \partial_{t^0} + \epsilon \partial_{t^1}.
\end{align*}
The term ``medium'' timescale is used to contrast with the fastest timescale, which is the
cyclotron frequency timescale.
The parallel total derivative at the medium timescale is defined as
\begin{align*}
    \dldtz = \partial_{t^0} + \Dl.
\end{align*}

\subsection{Order $\epsilon^{-1}$ kinetic equation}
Substituting \eqref{eqn:hilbert-expansion-f} into \eqref{eqn:vlasov_normalized} and
retaining only the leading-order term gives the order $\epsilon^{-1}$ kinetic equation
\begin{align}
    \label{eqn:kinetic-order-eps-inv}
    \frac{Z}{A}\oct \vp \times \bm{B} \cdot \nabla_{\bm{v}} f^0 = 0.
\end{align}
This can be rewritten as a homogeneous ordinary differential equation in the azimuthal (gyrophase) coordinate $\phi$,
defined via
\begin{align*}
\vp = (v_\perp \cos \phi, v_\perp \sin \phi)^T,
\end{align*}
and the species cyclotron frequency $\Omega_c = \omega_c \tau \frac{Z|B|}{A}$.
In terms of the azimuthal coordinate the leading-order kinetic equation is
\begin{align*}
-\Omega_c \partial_\phi f^0 = 0.
\end{align*}
Equation \eqref{eqn:kinetic-order-eps-inv} has general solutions that are gyrotropic in $\vp$,
that is, functions of $\vl$ and $|\vp|^2$.
However, in this work we assume that the leading-order solution is a Maxwellian,
\begin{align}
f^0 = \mathcal{M} = \frac{A\np (\bm{x}, \vl, t)}{2\pi \Tp^0 (\bm{x}, \vl, t)} \exp \left( -\frac{A |\vp|^2}{2\Tp^0(\bm{x}, \vl, t)} \right).
\end{align}
It is important to recognize that this is a modeling assumption.
It may be justified to assume that $f^0$ is a gyrotropic Maxwellian in certain cases,
particularly when flow velocities are not too large relative to the thermal velocity $v_{th}$.
In either case, the assumption of a Maxwellian leading-order solution is a modeling assumption
which may or may not match any particular physical situation.
We note that this assumption is not without precedent; for example, Ref. \onlinecite{simakovMomentumTransportArbitrary2007}
makes the same assumption on the leading-order solution.

Since any gyrotropic function of $\vp$ will satisfy \eqref{eqn:kinetic-order-eps-inv}, we are free
to choose its parameters.
We therefore assume that the density of $\mathcal{M}$ is equal to $\np$, implying that $\np^1 = \np^2 = \cdots = 0$.
Similarly, we assume that the total perpendicular energy of $\mathcal{M}$ is equal to $\ep$:
\begin{align*}
    \ep \triangleq \frac{A}{2} \int |\vp^2| f \dvp = \frac{A}{2} \int |\vp^2| \mathcal{M} \dvp.
\end{align*}
This implies $\ep = \pp^0 = \np \Tp^0$, and
higher-order corrections to the temperature and scalar pressure will
appear in subsequent equations.

The following expression for the perpendicular gradient of $\mathcal{M}$ will be useful later
\begin{align}
\Dp \mathcal{M} = \bm{p} \mathcal{M} + \frac{A |\vp|^2}{\Tp^0} \bm{r} \mathcal{M}, \quad \bm{p} = \frac{\Dp \np}{\np} - \frac{\Dp \Tp^0}{\Tp}, \quad \bm{r} = \frac{\Dp \Tp^0}{2\Tp^0}.
\end{align}

\subsection{Order $\epsilon^0$ kinetic equation}

Before proceeding to the order $\epsilon^0$ equation, we manipulate the governing
kinetic equation in such a way as to locate all of the solution momentum in $f^1$.
This is accomplished by adding the following equation to \eqref{eqn:vlasov_normalized}:
\begin{align*}
    -\Omega_c \frac{A}{\Tp^0} \partial_\phi (\epsilon^{-1} \up - \up^1 - \epsilon \up^2 - \cdots) \cdot \vp \mathcal{M} = 0.
\end{align*}
Here we have used the fact that $\up^0 = 0$.
Introduce the rescaled drift velocity $\ups = \epsilon^{-1} \up$, which is of order unity,
and collect order unity terms to obtain
\begin{align}
    \label{eqn:f1-kinetic}
    -\Omega_c \partial_\phi f^1 = -\dldtz \mathcal{M} - \vp \cdot \Dp \mathcal{M} - \opt\bm{E} \cdot \nabla_{\vp} \mathcal{M} + \npt C^0(f) - \Omega_c \frac{A}{\Tp^0} \partial_\phi (\ups - \up^1) \cdot \vp \mathcal{M}.
\end{align}
There is a Fredholm solvability condition on the right-hand side of \eqref{eqn:f1-kinetic} which is
that its gyroaverage vanish.
We introduce the following notation to split an arbitrary quantity $g$ into its gyro-averaged
component and the remainder:
\begin{align*}
    g = \overline{g} + \tilde{g} = \frac{1}{2\pi} \int_0^{2\pi} g(\phi^\prime) \mathrm{d} \phi^\prime + \left( g - \frac{1}{2\pi} \int_0^{2\pi} g(\phi^\prime) \mathrm{d}\phi^\prime \right).
\end{align*}
The Fredholm condition on \eqref{eqn:f1-kinetic} is therefore
\begin{align}
    \label{eqn:f1-fredholm}
    \dldtz \mathcal{M} = \npt \overline{C^0(f)}.
\end{align}
The remaining agyrotropic portion of \eqref{eqn:f1-kinetic} is
\begin{align}
    \label{eqn:f1-kinetic-agyrotropic}
\begin{split}
    -\Omega_c \partial_\phi f^1 &= -\vp \cdot \Dp \mathcal{M} - \bm{E} \cdot \nabla_{\vp} \mathcal{M} + \widetilde{C^0(f)}
    - \Omega_c \frac{A}{\Tp^0} \partial_\phi (\ups - \up^1) \cdot \vp \mathcal{M} \\
                                &= -\vp \cdot \left[ \bm{p} + \frac{A |\vp|^2}{\Tp^0} \bm{r} \right] \mathcal{M} + \frac{Z \opt \bm{E} \cdot \vp}{\Tp^0}
                                - \Omega_c \frac{A}{\Tp^0} \partial_\phi (\ups - \up^1) \cdot \vp \mathcal{M}.
\end{split}
\end{align}
We have neglected the agyrotropic component of $C^0(f)$, which vanishes for physically plausible collision operators:
since drift velocities are of order $\epsilon$, by a symmetry argument there is no mechanism for collisions to contribute agyrotropy
at leading order.
It is not hard to show that for a vector $\bm{g}$ independent of $\phi$, that
\begin{align*}
\int \bm{v} \cdot \bm{g} \,\mathrm{d} \phi = \bm{g} \cdot \int \bm{v} \mathrm{d} \phi = \frac{\bm{v} \cdot (\bm{B} \times \bm{g})}{|B|}.
\end{align*}
Thus, integrating \eqref{eqn:f1-kinetic-agyrotropic} in $\phi$, we get
\begin{align}
    \label{eqn:f1-partway}
    f^1 &= \frac{\vp}{\Omega_c |B|} \cdot \bm{B} \times \left[ \bm{p} - \frac{Z \opt \bm{E}}{\Tp^0} \right] \mathcal{M} + \frac{A |\vp|^2}{\Omega_c |B| \Tp^0} \vp \cdot (\bm{B} \times \bm{r}) + \frac{A(\ups - \up^1)}{\Tp^0} \cdot \vp \mathcal{M} \nonumber \\
        &= \frac{\vp}{\Omega_c |B|} \cdot \left[ \frac{\bm{B} \times \Dp \pp^0}{\pp^0} + \frac{Z\opt \bm{E} \times \bm{B}}{\Tp^0} \right] \mathcal{M} - 2 \frac{\vp \cdot \bm{B} \times \Dp \Tp^0}{\Omega_c |B| \Tp^0} \mathcal{M} + \frac{A |\vp|^2 \vp}{2 \Omega_c |B|} \cdot \frac{\bm{B} \times \Dp \Tp^0}{(\Tp^0)^2} \mathcal{M} \\
        &\qquad + \frac{A(\ups - \up^1)}{\Tp^0} \cdot \vp \mathcal{M} \nonumber
\end{align}
The first-order drift velocity $\up^1$ can be solved for by retaining terms of order unity in the
semi-fluid momentum equation \eqref{eqn:semifluid-momentum}:
\begin{align*}
\Dp \cdot \mathbb{P}^0 = \np Z (\opt \bm{E} + \oct \epsilon^{-1} \up^1 \times \bm{B}).
\end{align*}
We have neglected $\mathcal{S}^0$ by the same argument as above, namely that $C^0(f)$ is gyrotropic.
As can be verified by direct integration of $\mathcal{M}$, $\mathbb{P}^0 = \pp^0 \mathbb{I}$, 
so the first-order drift velocity is simply equal to the sum of the diamagnetic and $E \times B$ drifts:
\begin{align}
    \label{eqn:up1}
\up^1 = \frac{\bm{B} \times \Dp \pp^0}{\np Z \oct |B|^2} + \frac{\opt \bm{E} \times \bm{B}}{\oct |B|^2}.
\end{align}
Substituting \eqref{eqn:up1} into \eqref{eqn:f1-partway} and simplifying gives
\begin{align}
f^1 = \frac{A}{\Tp^0} (\ups - 2\bm{u}_T) \cdot \vp \mathcal{M} + \frac{A^2}{(\Tp^0)^2} \frac{\bm{u}_T}{2} \cdot \vp |\vp|^2 \mathcal{M},
\end{align}
where we have introduced the quantity $\bm{u}_T$ which has dimensions of velocity and is given by
\begin{align}
\bm{u}_T = \frac{\bm{B} \times \Dp \Tp^0}{Z \oct |B|^2}.
\end{align}
Using properties of the Maxwellian it is easy to verify that
\begin{align*}
\int \vp f^1 \dvp = \ups.
\end{align*}
The leading-order heat flux can now be computed as
\begin{align}
    \label{eqn:qp1-calc}
    \begin{split}
    \qp^1 &= -\frac{A}{2} \up \int |\vp|^2 \mathcal{M} \dvp - \frac{A}{2} \int 2(\vp \otimes \vp) \cdot \up \mathcal{M} \dvp + \frac{A}{2} \int |\vp|^2 \vp f^1 \dvp \\
          &= 2\pp^0 \bm{u}_T.
    \end{split}
\end{align}

\subsection{Order $\epsilon$ kinetic equation}
The kinetic equation at order $\epsilon$ is
\begin{align}
    \label{eqn:f2-kinetic}
    -\Omega_c \partial_\phi f^2 = -\partial_{t^1} \mathcal{M} - \dldtz f^1 - \vp \cdot \Dp f^1 - \frac{Z}{A}\opt \bm{E} \cdot \nabla_{\vp} f^1 + C^1(f) + \Omega_c \partial_\phi \frac{A}{\Tp^0} \up^2 \cdot \vp \mathcal{M}.
\end{align}
As before there is a Fredholm condition on the right-hand side.
It is up to the $\partial_{t^1}$ term to eliminate the gyrotropic components of the other terms
on the right-hand side.
To see how this occurs, we take the zeroth and second $\vp$ moments of \eqref{eqn:f2-kinetic} to obtain
the semi-fluid equations satisfied by $\np$ and $\Tp^0$:
\begin{align}
    \label{eqn:chapman-enskog-n}
    &\partial_{t^1} \np + \Dp \cdot (\np \ups) = \mathcal{N}^1 \\
    \label{eqn:chapman-enskog-p}
    &\partial_{t^1} \pp^0 + \Dp \cdot (2\pp^0 (\ups + \bm{u}_T)) = \np Z \opt \bm{E} \cdot \ups + \mathcal{Q}^1.
\end{align}
Equations \eqref{eqn:chapman-enskog-n} and \eqref{eqn:chapman-enskog-p} can be used to eliminate the $\partial_{t^1} \mathcal{M}$ term.
After expanding all gyrotropic terms in \eqref{eqn:f2-kinetic} and simplifying, we find that the
Fredholm condition reduces to 
\begin{align}
\label{eqn:fredholm-f2}
    \begin{split}
    0 &= \overline{
        \partial_{t^1} \mathcal{M} + \vp \cdot \Dp f^1 + \frac{Z}{A}\opt \bm{E} \cdot \nabla_{\vp} f^1 + C^1(f)
    } \\
      &= 
      \left( 2 \frac{\mathcal{N}^1}{\np} - \frac{\mathcal{Q}^1}{\pp^0} + \frac{A |\vp|^2}{2\Tp^0} \left( \frac{\mathcal{Q}^1}{\pp^0} - \frac{\mathcal{N}^1}{\np} \right)  \right) \mathcal{M} + \overline{C^1(f)} + \mathcal{O}(\epsilon).
    \end{split}
\end{align}
The gyrotropic terms will therefore vanish if $\overline{C^1(f)}$ vanishes,
in which case the gyrotropic moments $\mathcal{N}^1$ and $\mathcal{Q}^1$ also vanish.
In Section \ref{sec:landau-agyrotropic} we show that this condition holds
for the Landau-Fokker-Planck collision operator.

Subtracting the Fredholm condition from \eqref{eqn:f2-kinetic} gives the following equation for $f^2$:
\begin{align*}
    -\Omega_c \partial_\phi f^2 = -\widetilde{\vp \cdot \Dp f^1} - \frac{Z}{A} \opt \widetilde{\bm{E} \cdot \nabla_{\bm{v}} f^1} - \dldtz f^1 - \npt \widetilde{C^1(f)} + \Omega_c \partial_\phi \frac{A}{\Tp^0} \up^2 \cdot \vp \mathcal{M}.
\end{align*}
We can solve for $\up^2$ from the semi-fluid momentum equation.
Indeed, collecting terms of \eqref{eqn:semifluid-momentum} at order $\epsilon$, we find
\begin{align*}
A \dldtz (\np \up^1) + \Dp \cdot \mathbb{P}_\perp^1 = \np Z \oct \up^2 \times \bm{B} + \npt \mathcal{S}^1.
\end{align*}
As can be seen from the form of $f^1$ which has odd-order polynomial dependence on $\vp$, the first-order
pressure tensor $\mathbb{P}^1_\perp$ vanishes.
Solving for $\up^2$ gives
\begin{align*}
\up^2 = \frac{A \bm{B} \times \dldtz (\np \up^1)}{\np Z \oct |B|^2} + \frac{\npt \mathcal{S}^1 \times \bm{B}}{\np Z \oct |B|^2}.
\end{align*}
The first term is a polarization drift associated with the species inertia, while the second is a drift
produced by the collisional drag force $\mathcal{S}^1$.

\subsection{Collisionless gyroviscous stress}

We now calculate the leading-order gyroviscous stress in the collisionless limit and assuming symmetry
in the parallel direction.
In this limit the second-order drift velocity $\up^2$ vanishes, so the order $\epsilon$ kinetic equation
simplifies to
\begin{align*}
    -\Omega_c \partial_\phi f^2 &= -\widetilde{\vp \cdot \Dp f^1} - \frac{Z}{A}\opt \widetilde{\bm{E} \cdot \Dp f^1} \\
    &\triangleq RHS.
\end{align*}
The leading-order stress tensor contains a contribution from
\begin{align*}
A \int \vp \otimes \vp f^2 \dvp = -\frac{1}{\Omega_c} \int \vp \otimes \vp (RHS) \mathrm{d} \phi.
\end{align*}
It can be shown without solving for $f^2$ that 
\begin{align}
    \label{eqn:second-moment-f2}
    A \int \vp \otimes \vp f^2 \dvp = \pp^0 \frac{\mathbb{W}_3 [\ups]}{2\Omega_c} + \frac{\mathbb{W}_3[\bm{q}^1_\perp]}{4\Omega_c} + \frac{A}{\Tp^0} \left( \frac{\pp^0 \widehat{(\up^1 \otimes \ups)}}{2} + \frac{\widehat{(\bm{q}_\perp^1 \otimes \ups)}}{4} + \frac{(\bm{q}^1_\perp \otimes \bm{q}^1_\perp)}{2\pp^0} \right),
\end{align}
where the symmetric trace-free tensor $\mathbb{W}_3[\bm{a}]$ is defined for an arbitrary vector $\bm{a} = (a_x, a_y)$ via
\begin{align*}
    \mathbb{W}_3[\bm{a}] = \epsilon_{xy \parallel} \begin{pmatrix}
        -\partial_y a_x - \partial_x a_y & \partial_x a_x - \partial_y a_y \\
        \partial_x a_x - \partial_y a_y & \partial_y a_x + \partial_x a_y
    \end{pmatrix},
\end{align*}
with $\epsilon_{xy\parallel}$ a Levi-Civita symbol indicating the orientation of the triplet $(x, y, \parallel)$
for perpendicular coordinates $x, y$.
The notation $\widehat{\mathbb{A}}$ is defined as the trace-free symmetrization of a $2 \times 2$ tensor $\mathbb{A}$,
\begin{align}
    \label{eqn:defn-widehat}
    \widehat{\mathbb{A}} = \mathbb{A} + \mathbb{A}^T - \text{Tr}(\mathbb{A}) \mathbb{I}.
\end{align}
Subtracting the scalar pressure $p$ from the pressure tensor gives us our closure expression for $\Pi$:
\begin{align}
    \label{eqn:Pi_perp}
    \begin{split}
    \Pi_\perp &= \mathbb{P}_\perp - p \mathbb{I} \\
        &= \mathbb{P}_\perp - \left( \pp^0 - \frac{A\np |\up|^2}{2} \right) \mathbb{I} \\
        &= \pp^0 \frac{\mathbb{W}_3 [\ups]}{2\Omega_c} + \frac{\mathbb{W}_3[\qp^1]}{4\Omega_c} + 
        \frac{A\np}{2} [\widehat{(\up^\dagger - \ups) \otimes \ups}] + \frac{A}{2\pp^0 \Tp^0} \widehat{(\qp^1 \otimes \qp^1)}.
    \end{split}
\end{align}
We have simplified the expression slightly by using \eqref{eqn:qp1-calc} and introducing the velocity $\up^\dagger = \up^1 + \bm{u}_T$.

\subsection{Summary of semi-fluid equations with leading-order FLR effects}

We briefly summarize the set of closed semi-fluid equations derived above which incorporate
FLR effects to leading order.
They are
\begin{align}
    \label{eqn:semifluid-np}
&\dldt \np + \Dp \cdot (\np \up) = \npt \mathcal{N} \\
    \label{eqn:semifluid-unp}
    A &\dldt (\np \up) + \Dp \cdot (A \np \up \otimes \up + \pp\mathbb{I} + \Pi_\perp) = \np Z (\opt \bm{E} + \oct \up \times \bm{B}) + \mathcal{S} \\
    \label{eqn:semifluid-ep}
      &\dldt (\ep) + \Dp \cdot ((e \mathbb{I} + \pp \mathbb{I} + \Pi_\perp) \cdot \up + \qp) = \np Z \opt \up \cdot \bm{E} + \mathcal{Q}.
\end{align}
The collisional moments $\mathcal{N}, \mathcal{S}, \mathcal{Q}$ can be calculated by taking
moments of a specific collision operator, expanded to second-order in $\epsilon$.
For a bilinear collision operator such as the Landau operator, such an expansion is straightforward:
\begin{align*}
    C(f_s, f_\sp) = C(f_s^0, f_\sp^0) + \epsilon \left[ C(f_s^1, f_\sp^0) + C(f_s^0, f_\sp^1) \right] + \dots
\end{align*}
The stress tensor $\Pi_\perp$ appearing in \eqref{eqn:semifluid-unp} and \eqref{eqn:semifluid-ep} is given by \eqref{eqn:Pi_perp}, while the heat flux appearing in \eqref{eqn:semifluid-ep} is given by
\eqref{eqn:qp1-calc}.

\subsection{Drift-advection limit of electron momentum}
\label{sec:drift-advection}

The derivation leading to the semi-fluid equations \eqref{eqn:semifluid-np}-\eqref{eqn:semifluid-ep}
has been agnostic of the magnitude of the species mass, and thus has equal formal accuracy for
electrons and ions.
However, particularly in the low-beta regime considered here, the electron fluid model can be greatly
simplified by observing that their inertia is negligible.
Indeed, taking the limit of $A = A_e \rightarrow 0$, we find that $\Pi_\perp$ 
vanishes since it is proportional to $A$ through the inverse species cyclotron frequency
$\Omega_c^{-1}$.
Neglecting the inertial terms in the electron semi-fluid momentum equation, then, we obtain
\begin{align}
    \label{eqn:semifluid-generalized-ohms}
\Dp \cdot (\ppe \mathbb{I}) = \npe Z_e (\opt \bm{E} + \oct \upe \times \bm{B}),
\end{align}
which can be solved to verify that the electron velocity is equal to the sum of the 
electron diamagnetic and $E \times B$ drifts.
Equation \eqref{eqn:semifluid-generalized-ohms} is the generalized Ohm's law for the
semi-fluid system of equations.
Note that it is missing a source term on the right-hand side which would
account for interspecies collisions, i.e. resistivity.
This reflects the fact that in the ordering chosen here, collisions are ordered weaker
than magnetic forces, and thus resistivity is neglected at leading order.

Substituting \eqref{eqn:semifluid-generalized-ohms} result into the continuity equation, and using the fact that
the diamagnetic current is divergence free, we obtain the greatly simplified
electron continuity equation,
\begin{align}
    \label{eqn:electron-ExB}
\partial_t \npe + \Dp \cdot \left( \npe \frac{\opt \bm{E} \times \bm{B}}{\oct |B|^2} \right) = 0.
\end{align}
The electron continuity equation is therefore seen to be independent of the electron
energy equation, which is
\begin{align}
    \label{eqn:electron-energy}
\partial_t \epe + \Dp \cdot \left( (\epe + \ppe) \upe + \qpe \right) = \npe Z_e \opt \upe \cdot \bm{E}.
\end{align}
Note that we cannot say that the electron energy equation \eqref{eqn:electron-energy} is independent
of the continuity equation \eqref{eqn:electron-ExB} by virtue of the density-dependent work term
on the right-hand side.
The lone equation \eqref{eqn:electron-ExB} is an asymptotically consistent model
of electron motion in the low-beta, small mass ratio limit.
The virtue of \eqref{eqn:electron-ExB} from a numerical point of view is that it does not resolve
electron plasma oscillations which can impose highly restrictive maximum timestep constraints
on explicit solvers.
When incorporated into a two-fluid model however, \eqref{eqn:electron-ExB} retains
essential charge separation physics, making it a useful approximation for two-fluid plasmas
which evolve on ion timescales.

\subsection{Approximate and numerically feasible gyroviscous stress tensor}
\label{sec:numerically-feasible-adjustment}

A key benefit of the Braginskii closure is that the first-order transport terms that it predicts
are diffusive, meaning that they involve second-order derivatives of primary quantities
such as density, velocity, and temperature.
The same cannot be said for \eqref{eqn:Pi_perp} which, when substituted into \eqref{eqn:semifluid-unp},
introduces a third-order derivative of temperature on the right-hand side of the momentum equation.
This is highly inconvenient for numerical implementation since it implies eigenvalues that
may grow as $\Delta x^{-3}$ for a discretization with grid scale $\Delta x$.
For explicit time discretizations, the maximum stable timestep decreases as $\Delta x^3$,
compared to $\Delta x^2$ for a diffusive equation.
We therefore seek an approximation to the gyroviscous stress tensor which takes into account
the temperature gradient effects contained in \eqref{eqn:Pi_perp} without imposing a $\Delta x^3$
stability requirement for explicit schemes.

For drift-dominated flows, this can be achieved with a simple modification 
to the Braginskii gyroviscous stress.
To proceed, we use the fact that the fluid velocity is small relative to the thermal
velocity, which is implied by our leading-order gyrotropy assumption.
Neglecting terms in \eqref{eqn:Pi_perp} which are quadratic in a fluid velocity, and 
using \eqref{eqn:qp1-calc}, we find
\begin{align*}
    \Pi \approx \frac{\pp^0}{2\Omega_c} \mathbb{W}_3 \left[ \bm{u}_E + \bm{u}_d + \bm{u}_T \right],
\end{align*}
where $\bm{u}_E$ and $\bm{u}_d$ are the $E \times B$ and diamagnetic velocity respectively:
\begin{align*}
\bm{u}_E = \frac{\opt \bm{E} \times \bm{B}}{\oct |B|^2}, 
\quad 
\bm{u}_d = \frac{\bm{B} \times \Dp \pp}{\np Z \oct |B|^2}.
\end{align*}
The Braginskii gyroviscous stress closure is
\begin{align}
    \label{eqn:Pi_Brag}
    \Pi^{Brag} \approx \frac{\pp}{2\Omega_c} \mathbb{W}_3 \left[ \bm{u}_E + \bm{u}_d \right].
\end{align}
In the presence of temperature gradients, the discrepancy between
the Braginskii and drift-ordering gyroviscous stresses can be estimated using the factor
\begin{align*}
    \gamma^{Brag} \triangleq \frac{|\nabla p / p + \nabla T / T|}{|\nabla p / p|} \approx \frac{|\bm{u}_d + \bm{u}_T|}{|\bm{u}_d|}.
\end{align*}
Therefore, we expect the following adjustment to the Braginskii gyroviscous stress to be a
good approximation to \eqref{eqn:Pi_perp}:
\begin{align}
    \label{eqn:Pi_adj}
    \Pi^{Adj} = \frac{\pp}{2\Omega_c} \left( \mathbb{W}_3[\bm{u}_E] + \gamma^{Brag} \mathbb{W}_3 [\up - \bm{u}_E] \right).
\end{align}
Note that \eqref{eqn:Pi_adj} does not require calculating either $\bm{u}_d$ or $\bm{u}_T$;
rather, it relies on the assumption that the fluid velocity is dominantly composed of the $E \times B$
and diamagnetic velocity.
We have also used the fact that $\pp = \pp^0 + \mathcal{O}(\epsilon^2)$.

The factor $\gamma^{Brag}$ can be set either as a global simulation parameter or determined
locally from estimates of the local gradient scale lengths.
In the simulations reported here, we use a global estimate of $\gamma^{Brag}$.

\section{Kinetic simulation}
\label{sec:kinetic_modeling}

Our numerical experiments are conducted using a high-accuracy continuum kinetic solver
for the Vlasov equation in two perpendicular dimensions (``2D2V'').
Continuum kinetic simulation is a still-emerging methodology which offers significant advantages
for investigating the detailed structure of solutions to the Vlasov equation.
A key benefit of continuum kinetic simulation is that it provides a solution for the full kinetic
distribution function.
This allows one to compute kinetic values for the closure
moments---the heat flux and stress tensor---and compare
them to the leading-order transport closures derived in the previous section.

The simulations conducted in this paper use a hybrid simulation approach which couples fully
kinetic ions to fluid electrons.
By representing the ion species distribution function explicitly, the code captures ion
finite Larmor radius effects with high fidelity.
The electron species is solved with the drift-advection continuity equation derived in 
Section \ref{sec:drift-advection}.
As is justified by the low-beta regime, we use the electrostatic approximation, which neglects
plasma current contributions to the magnetic field and solves Gauss's law for the electrostatic potential.
We assume negligible collisions ($\npt = 0$), as well as assuming symmetry in the parallel direction
($\mathcal{D}_\parallel = 0$).

To summarize, the governing equations solved by the hybrid kinetic-fluid code are
\begin{align}
    \label{eqn:ion-vlasov}
    &\partial_t f_i + \vp \cdot \Dp f_i + \frac{Z_i}{A_i} \left( \opt \bm{E}_\perp + \oct \vp \times \bm{B}_0 \right)  \cdot \nabla_{\vp} f_i = 0 \\
    \label{eqn:electron-drift-advection}
    &\partial_t n_e + \Dp \cdot (n_e \bm{u}_E) = 0 \\
    \label{eqn:gauss-law}
    &-\nabla^2 \phi = \opt \sum_{s=i,e} n_s Z_s,
\end{align}
where the $E \times B$ velocity is defined by
\begin{align*}
\bm{u}_E = \frac{\opt \bm{E} \times \bm{B}_0}{\oct |B_0|^2}.
\end{align*}
Equations \eqref{eqn:ion-vlasov}-\eqref{eqn:gauss-law} are solved in a two-dimensional spatial domain
with the magnetic field in the direction of symmetry.
We choose the numerical coordinate system so that $\bm{B} = B_0 \hat{\bm{y}}$, thus the perpendicular
coordinates are labeled $x$ and $z$.
The $x$ dimension is equipped with periodic boundary conditions while the $z$ dimension has a finite width.
In the $z$ direction, we employ a ``reservoir'' boundary condition for the Vlasov
equation \cite{cagasBoundaryValueReservoir2021, rodmanKineticInterpretationClassical2022}
which uses ghost cells that are set to a continuation of the initial condition beyond the boundary.
Boundary conditions on the electric potential for Gauss's law are constant in time and equal to the initial condition.

The Fourier-Hermite discretization used to solve equations \eqref{eqn:ion-vlasov} and 
\eqref{eqn:electron-drift-advection} is described in Section \eqref{eqn:numerical_methods}.
A Hermite spectral discretization is particularly advantageous for problems in the slow-dynamics
regime we address here, where fluid velocities are lower than thermal velocities.
For this reason, we observe good accuracy with as few as $N_{v_x} = N_{v_z} = 26$ Hermite
modes in each velocity dimension.
All problems are solved with $N_x = 144$ Fourier modes in the $x$ direction and $N_z = 280$ grid
points in $z$.

\subsection{Kinetic initial condition}

In this section we describe a family of kinetic initial conditions which include vorticity,
sheared flow, and temperature gradients.
These plasma configurations resemble the magnetized Kelvin-Helmholtz instability at late
times in the linear phase.
In order to study physics related to non-equilibrium transport, however, we initialize
a plasma that is far from equilibrium, rather than the equilibrium initialization that is
typical of studies of fluid instabilities.
To minimize the impact of transient waves on the solution, we construct initial conditions satisfying
the equation
\begin{align}
    \label{eqn:unchanging}
    \dtt \left. \begin{pmatrix}
    n_i \\ n_i \bm{u}_i \\ T_i
    \end{pmatrix}
    \right\vert_{t=0} = 0.
\end{align}
This condition results in a clean initialization to the simulation with no compressive waves,
which tend to oscillate on timescales which are fast relative to the bulk motion of the plasma
and complicate interpretation of the solution.

The non-conservative form of the five-moment equations is
\begin{align}
\label{eqn:nonconservative-5m}
\begin{split}
    &\dtt n_i + n_i \nabla \cdot \bm{u}_i = 0, \\
    A_i & \dtt (n_i \bm{u}_i) + \nabla p_i = n_i Z_i (\opt \bm{E} + \oct \bm{u}_i \times \bm{B}), \\
      &\dtt T_i + (\gamma - 1) T_i \nabla \cdot \bm{u}_i = 0,
\end{split}
\end{align}
where $\gamma$ is the ratio of specific heats.
From \eqref{eqn:nonconservative-5m}, we see that \eqref{eqn:unchanging} will be satisfied
if $\nabla \cdot \bm{u}_i = 0$ and
\begin{align*}
\bm{u}_i = \frac{\opt \bm{E} \times \bm{B}}{\oct |B|^2} + \frac{\bm{B} \times \nabla p_i}{n_i Z_i |B|^2}.
\end{align*}
The incompressibility condition is automatically satisfied by the $E \times B$ drift.
For the diamagnetic drift we calculate
\begin{align}
    \nabla \cdot \left( \frac{\bm{B} \times \nabla p_i}{n_i} \right) &= \nabla \cdot \left( \bm{B} \times \nabla T_i + T_i(\bm{B} \times \nabla (\ln n_i)) \right) \\
                                                                 &= \nabla T_i \cdot (\bm{B} \times \nabla (\ln n_i)).
\end{align}
Thus, the ion diamagnetic velocity will be incompressible as long as $\nabla n_i$ and $\nabla T_i$ are colinear,
corresponding to no Biermann battery effect.
This consideration motivates us to base the initial condition on an overall ion pressure profile
function $\hat{p}_i(z)$ defined by
\begin{align*}
\hat{p}_i(z) = 1 + \gamma \tanh \left( \frac{z}{\alpha} \right),
\end{align*}
where $\gamma$ and $\alpha$ are parameters setting the magnitude and width of the interface jump, respectively.
We control the relative variation of ion density and temperature via a parameter $\zeta$:
\begin{align}
    \label{eqn:ni0-Ti0}
    n_{i0}(z) = n_{ref} \hat{p}(z)^\zeta, \quad T_{i0}(z) = T_{ref} \hat{p}(z)^{1-\zeta}.
\end{align}

In addition to a pressure gradient, we initialize an $E \times B$ flow field with both shear
and vorticity, controlled by the parameters $u_s$ and $u_V$ respectively.
The first, $u_s$, represents the desired change in $u_x$ from the bottom to the top of the domain:
\begin{align*}
    u_s \triangleq \left. \frac{-\opt E_z}{\oct B} \right\vert_{z=L_z/2} - \left. \frac{- \opt E_z}{\oct B} \right\vert_{z=-L_z/2}.
\end{align*}
The latter represents the desired maximum $z$-directed velocity at the center of the domain:
\begin{align*}
    u_V \triangleq \max_{x} \left. \frac{\opt E_x}{\oct B} \right\vert_{z=0}. 
\end{align*}
We obtain the desired $E \times B$ velocities by prescribing an electrostatic potential
$\phi^*(x, z)$ given by
\begin{align*}
\phi^*(x, z) = \frac{\oct B_0}{\opt} \phi^*_X(x, z) \phi^*_Z(z),
\end{align*}
where
\begin{align*}
    \phi^*_Z(z) = \left(1 + \frac{u_s \alpha}{2} \ln \cosh \left( \frac{z}{\alpha} \right)\right),
\end{align*}
\begin{align*}
    \phi^*_X(x, z) = 1 + \frac{u_V}{k_x} \sin(k_x x) \exp \left( -\frac{z^2}{2 w^2} \right).
\end{align*}
We have introduced two further geometric parameters, $k_x$ and $w$, which set the wavenumber and width, respectively,
of the vorticity.
Given $\phi^*$, the desired charge potential $\rho_c^*$ is taken to satisfy Gauss's law \eqref{eqn:gauss-law}, from which we can calculate the initial electron density via
\begin{align*}
    n_{e0} = \frac{1}{Z_e} \left(\rho_c^* - Z_i n_{i0}\right).
\end{align*}

\begin{figure}
    \begin{subfigure}[t]{0.5\textwidth}
        \includegraphics[width=\textwidth]{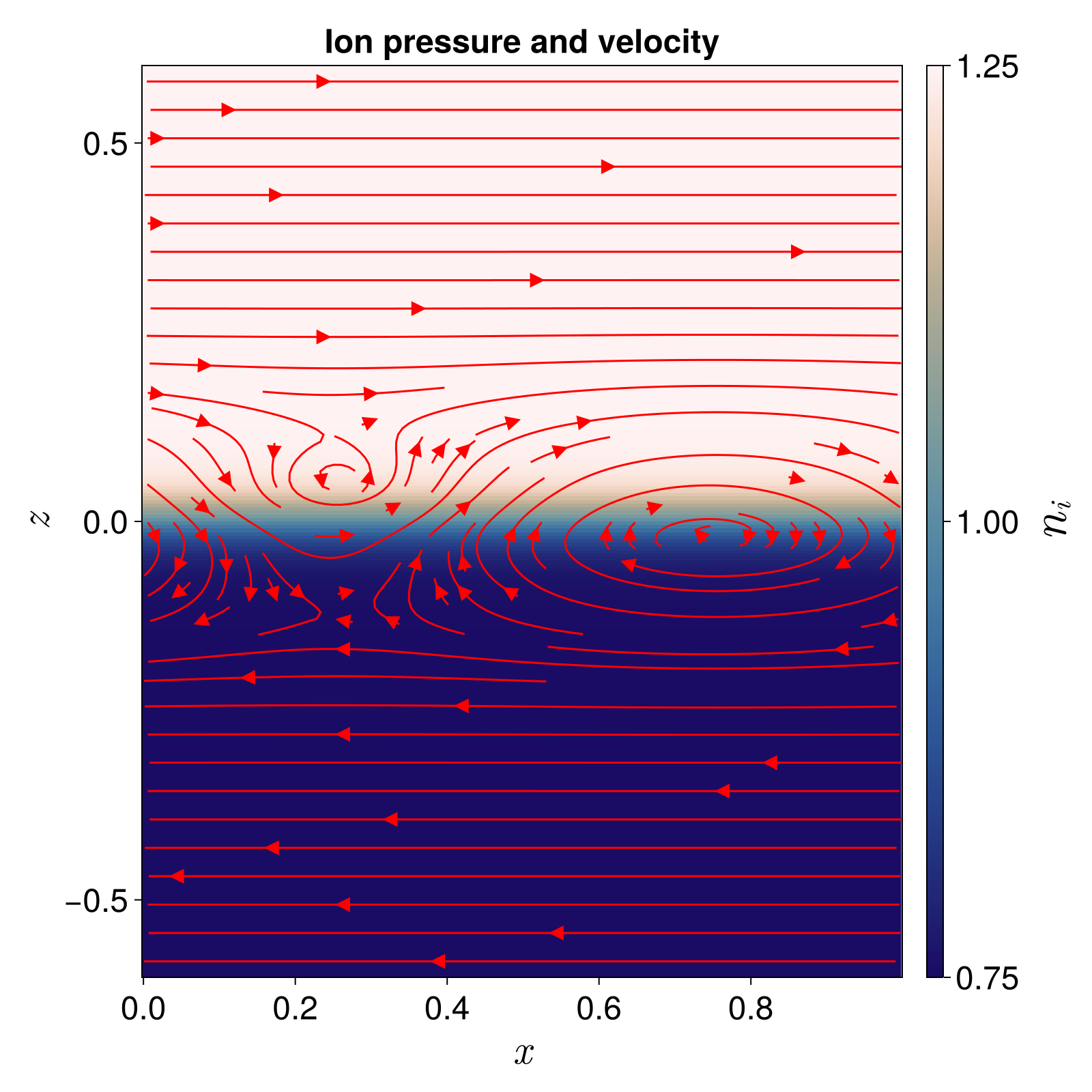}
        \caption{Plot of ion density $n_{i0}$ overlaid with ion velocity $\bm{u}_{i0}$.}
    \end{subfigure}
    \begin{subfigure}[t]{0.5\textwidth}
        \includegraphics[width=\textwidth]{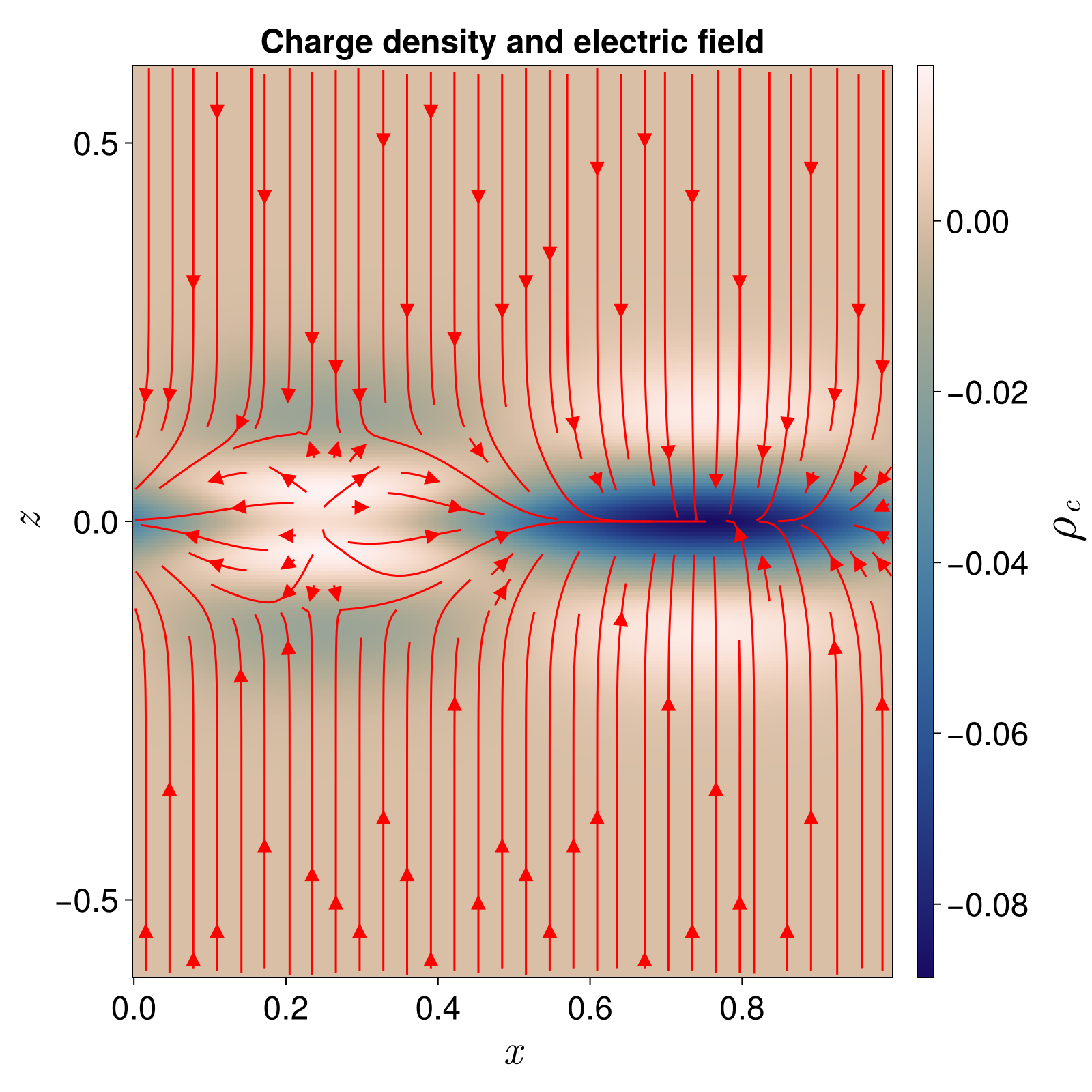}
        \caption{Plot of charge density $\rho_{c0}$ overlaid with electric field $\bm{E}_0$.}
    \end{subfigure}
    \caption{Illustrative example of an incompressible ion flow initial condition in a domain with sizes $L_x = 1.0,\ L_z = 1.2$. 
    The parameters chosen are $\gamma = 0.25,\ \omega_c \tau = \omega_p \tau = 2.0,\ u_s = 0.2v_{ti},\ u_V = 0.1v_{ti}$,
where $v_{ti} = \sqrt{T_{ref} / A_i}$ is the ion thermal speed. The geometric parameters are $\alpha = 0.04$ and $w = 2\alpha$. \label{fig:example-ic}}
\end{figure}

An example initial condition is plotted in Figure \ref{fig:example-ic}.

In addition to the incompressibility condition on the Maxwellian fluid variables \eqref{eqn:unchanging},
we also apply a non-Maxwellian initial condition to the pressure tensor and heat flux.
To mitigate the effects of waves on the higher moments of the solution, we initialize
the stress tensor and heat flux to their leading-order values as predicted by
\eqref{eqn:Pi_perp} and \eqref{eqn:qp1-calc}.
That is, we seek an ion initial condition $f_{i0}$ satisfying
\begin{align*}
\mathbb{P}_{i0} &= A_i \int (\bm{v} - \bm{u}_{i0}) \otimes (\bm{v} - \bm{u}_{i0}) f_{i0}\dv = p_{i0} \mathbb{I} + \left[ p_{i0} \frac{\mathbb{W}_3[\bm{u} + \bm{u}_{Ti0}]}{2\Omega_{ci}} + \frac{\mathbb{W}_3[\bm{q}_{i0}]}{4\Omega_{ci}} + \frac{A_i n_{i0}}{2} \widehat{\mathsf{U}} \right], \\
    \bm{q}_{i0} &= \frac{A_i}{2} \int (\bm{v} - \bm{u}_{i0}) |\bm{v} - \bm{u}_{i0}|^2 f_{i0} \dv =
    2p_{i0} \bm{u}_{Ti0},
\end{align*}
where the $\widehat{\cdot}$ notation indicates the trace-free symmetrization defined in \eqref{eqn:defn-widehat},
\begin{align*}
    \mathsf{U} = \bm{u}_{Ti0} \otimes (\bm{u}_{i0} + \bm{u}_{Ti0}),
\end{align*}
and
\begin{align*}
    \bm{u}_{Ti0} = \frac{\bm{B} \times \nabla T_{i0}}{Z \oct |B|^2}.
\end{align*}

The full pressure tensor is easily prescribed with a non-isotropic Maxwellian distribution:
\begin{align}
    \label{eqn:fhat-1}
    \hat{f}_i = \frac{A_i n_{i0}}{2\pi |\mathbb{T}_{i0}|^{1/2}} \exp \left( -\frac{A_i \bm{w}^T \mathbb{T}_{i0}^{-1} \bm{w}}{2} \right),
\end{align}
where $\bm{w} = \bm{v} - \bm{u}_{i0}$ is the relative velocity,
$\mathbb{T} = \frac{1}{n_{i0}} \mathbb{P}_{i0}$ is the temperature tensor, 
$\mathbb{T}^{-1}$ denotes the matrix inverse, and $|\mathbb{T}|$ the matrix determinant.
That \eqref{eqn:fhat-1} gives a distribution with the correct pressure tensor
can be verified using standard properties of the multivariate Gaussian distribution having
$\mathbb{T}$ as a covariance matrix.

To additionally prescribe the correct heat flux, we add a component to $\hat{f}_i$ having
heat flux $\bm{q}_{i0}$ and vanishing lower moments.
This can be accomplished by defining
\begin{align}
    \label{eqn:fi0}
    f_{i0} = \hat{f}_i + \mathcal{M}_{i0} \left[ \frac{2\bm{q}_{i0}}{A_i n_{i0} v_{ti}^3 \sqrt{6}} \cdot \bm{H}_3 \right],
\end{align}
where $v_{ti} = \sqrt{T_{i0}/A_i}$,
\begin{align*}
    \bm{H}_3 = \left( He_3 \left( \frac{w_x}{v_{ti}} \right), He_3 \left( \frac{w_z}{v_{ti}} \right)   \right)^T,
\end{align*}
and
\begin{align*}
    \mathcal{M}_{i0} = \frac{A_i n_{i0}}{2\pi T_{i0}} \exp \left( -\frac{A_i |\bm{w}|^2}{2 T_{i0}} \right) 
\end{align*}
is the local Maxwellian with parameters $n_{i0}, \bm{u}_{i0}, T_{i0}$.
It can be verified by direct integration and using orthogonality properties of the Hermite
polynomials that \eqref{eqn:fi0} has the desired density, velocity, pressure tensor, and heat flux.

To evaluate the regions of validity of our collisionless, magnetized transport theory,
we perform several simulations with varying parameter values.
The parameters are summarized in Table \ref{table:sf-simulation-table}.
Series A is designed to explore the role of magnetization in the validity of the leading-order
transport theory.
Magnetization is characterized by the dimensionless parameter $\oct$, which in the 
asymptotic expansion of Section \ref{sec:semi_fluid_closure} is formally connected to the 
small parameter $\epsilon$.
Series A consists of seven simulations with $\oct$ varying from $0.5$ to 4.5.
For reference, the plasma frequency is set by $\opt = 1$.
Thus, simulation A1 is weakly magnetized relative to electrostatic effects, while simulation
A7 is strongly magnetized.
Series A fixes the parameter $\zeta$ at 0.5, which balances the density and temperature
contributions to the pressure gradient (and therefore diamagnetic drift).
Thus, the heat flux correction to the gyroviscous stress tensor is expected to play a role
in these simulations.

Series B and C are designed to explore the role of temperature gradients in 
driving the gyroviscous stress.
In these series the parameter $\zeta$ is varied from -0.5 to 2.0 in increments of 0.5.
Per \eqref{eqn:ni0-Ti0}, a value of $\zeta = 0.0$ represents a uniform density
profile, while $\zeta = 1.0$ represents an isothermal initial condition.
Setting $\zeta = -0.5$ gives a large temperature gradient and a density profile 
which is oriented opposite the pressure gradient, while $\zeta = 2.0$ gives the
reverse: a large density gradient and a temperature gradient oriented
opposite the pressure gradient.
By varying the relative contribution of temperature to the pressure gradient,
we control the relative magnitude of the heat flux and diamagnetic drift in
the shear layer, and correspondingly, the relative magnitude of the first
two terms of \eqref{eqn:Pi_perp}.
A larger relative contribution of the second term of \eqref{eqn:Pi_perp}
corresponds to larger deviation from the Braginskii gyroviscous stress closure,
as discussed in Section \ref{sec:numerically-feasible-adjustment}.
In this way we can investigate the importance of temperature gradients in driving
gyroviscous transport of momentum.
Moreover, series B and C are run with different values of $\omega_c \tau$, with the
aim of elucidating the importance of magnetization on the heat flux correction.

To evaluate the role of nonlinear turbulent dynamics in transport closure validity,
we run a series of simulations with a superposition of multiple sinusoidal
modes in the initial velocity field, series M.
We generalize the imposed electrostatic potential by defining
\begin{align*}
    \phi_X^*(x, z) = 1 + \sum_{i} \frac{u_V}{k_x^i} \sin(k_x^i x) \exp \left( -\frac{z^2}{2w^2} \right),
\end{align*}
for a collection of wavenumbers $k_x^i$.
We apply two modes with wavenumbers $k_x^1 = \pi, k_x^2 = 2\pi$.
Additionally we widen the domain to $L_x = 2.0$, and reduce the vorticity velocity $u_V$ compared
to series A.

Finally, we seek to understand the role of the polarity of sheared flow in FLR effects.
This is accomplished through simulations S1 and S2, which are initialized with opposite
shear polarities, defined as the sign of $(\nabla \times \bm{u}) \cdot \bm{B}$.
The polarity of the sheared flow relative to the magnetic field has been found to impact
the linear growth rate of magnetized Kelvin-Helmholtz instabilities 
\cite{umedaIonKineticEffects2014, 
umedaEvaluatingGyroviscosityKelvinHelmholtz2016, 
vogmanTwofluidKineticTransport2020}.
This effect was observed in Ref. \onlinecite{vogmanTwofluidKineticTransport2020} to be connected
to ion inertia through the polarization drift.

\begin{table}
    \small
    \centering
    \begin{tabular}{c|l|c|c|c|c|c}
        & Description & A1-A7 & B1-B6 & C1-C6 & M1-M4 & S1-2 \\
        \hline
        \rowcolor{tablerowgray}
        $\omega_c \tau$ & Magnetization & $\{ 0.5, 0.75, \dots, 4.5 \}$ & 2.0 & 4.0 & $\{ 1.5, 2.0, 3.0, 4.5 \}$ & 2.0\\
        $\gamma$ & Pressure jump & \multicolumn{4}{c|}{$0.4$} & 0.25 \\
        \rowcolor{tablerowgray}
        $u_s$ & Shear velocity & \multicolumn{4}{c|}{$0.2v_{ti}$} & $\pm 0.2v_{ti}$ \\
        $u_V$ & Vortex velocity & \multicolumn{3}{c|}{$0.06v_{ti}$} & $0.036v_{ti}$ & $0.1v_{ti}$ \\
        \rowcolor{tablerowgray}
        $\zeta$ & Density/temperature balance & $0.5$ & \multicolumn{2}{|c|}{$\{-0.5, 0.0, \dots 2.0\}$} & 0.5 & 0.5 \\
        $k_x$ & Wavenumber & \multicolumn{3}{c|}{$2\pi$} & $\{ 2\pi, 4\pi \}$ & $2\pi$ \\
        \rowcolor{tablerowgray}
        $T_{ref}$ & Reference temperature & \multicolumn{5}{c}{$\num{1e-3}$} \\
        $\alpha$ & Interface width & \multicolumn{5}{c}{$0.04$} \\
        \rowcolor{tablerowgray}
        $A_e$ & Electron mass & \multicolumn{5}{c}{$1/1836$}
    \end{tabular}
    \caption{Summary of simulation parameter values.
    \label{table:sf-simulation-table}
    }
\end{table}

\section{Numerical results}
\label{sec:numerical_results}

\subsection{Leading-order convergence}

To summarize the predictive performance of the leading-order closures \eqref{eqn:qp1-calc} and
\eqref{eqn:Pi_perp}, we use a standard measure of ``goodness of fit'', namely the $R^2$ value
for a predictive model.
$R^2$ is defined as follows for a model variable $\hat{\psi}$ intended to approximate
the ground truth value $\psi$:
\begin{align}
    \label{eqn:sf-R2-def}
R^2 = 1 - \frac{|\psi - \hat{\psi}|^2}{|\psi - \overline{\psi}|^2},
\end{align}
where $\overline{\psi}$ is the average of $\psi$.
For the vector- and tensor-valued transport relations evaluated here, we compute the error and mean
componentwise, and then integrate to find the $L^2$ norm of the 
error and deviation from the mean. That is, we compute
\begin{align*}
    R^2_\Pi(t) &= 1 - \frac{\left(  \int_\Omega  \|\Pi_\perp(\bm{x}) - \widehat{\Pi}_\perp(\bm{x})\|_2^2 \dx  \right)^{1/2}}{\left(  \int_\Omega  \|\Pi_\perp(\bm{x}) - \overline{\Pi_\perp}\|_2^2 \dx  \right)^{1/2}}, \\
    R^2_{\bm{q}}(t) &= 1 - \frac{\left(  \int_\Omega  \|\bm{q}_\perp(\bm{x}) - \widehat{\bm{q}}_\perp(\bm{x})\|_2^2 \dx  \right)^{1/2}}{\left(  \int_\Omega  \|\bm{q}_\perp(\bm{x}) - \overline{\bm{q}_\perp}\|_2^2 \dx  \right)^{1/2}},
\end{align*}
where $\overline{\bm{q}}$ and $\overline{\Pi}$ are spatial average quantities and $\| \cdot \|_2$ denotes the $L^2$ norm.

Snapshots are taken of the kinetic simulations by taking a weighted average of $f$ 
over a time period of length $0.75 \tau$, which for the simulations performed here
ranges from approximately one cyclotron period to around 5 cyclotron periods.
These weighted averages are then processed by taking moments to obtain $\Pi_\perp$, $\bm{q}_\perp$,
and the inputs to the closures \eqref{eqn:Pi_perp} and \eqref{eqn:qp1-calc}.
The averaging process smooths over fast variations due to wave phenomena at close to
the cyclotron and plasma frequencies, while leaving
the long-time evolution of the moments and transport closures unaffected.
In order to better center the snapshots at a point in time, the weighting function is chosen
to be a ``hat'' function which is piecewise linear and symmetric about the point in time
to which the snapshot is attributed.

\begin{figure}
    \centering
    \includegraphics[width=0.5\textwidth]{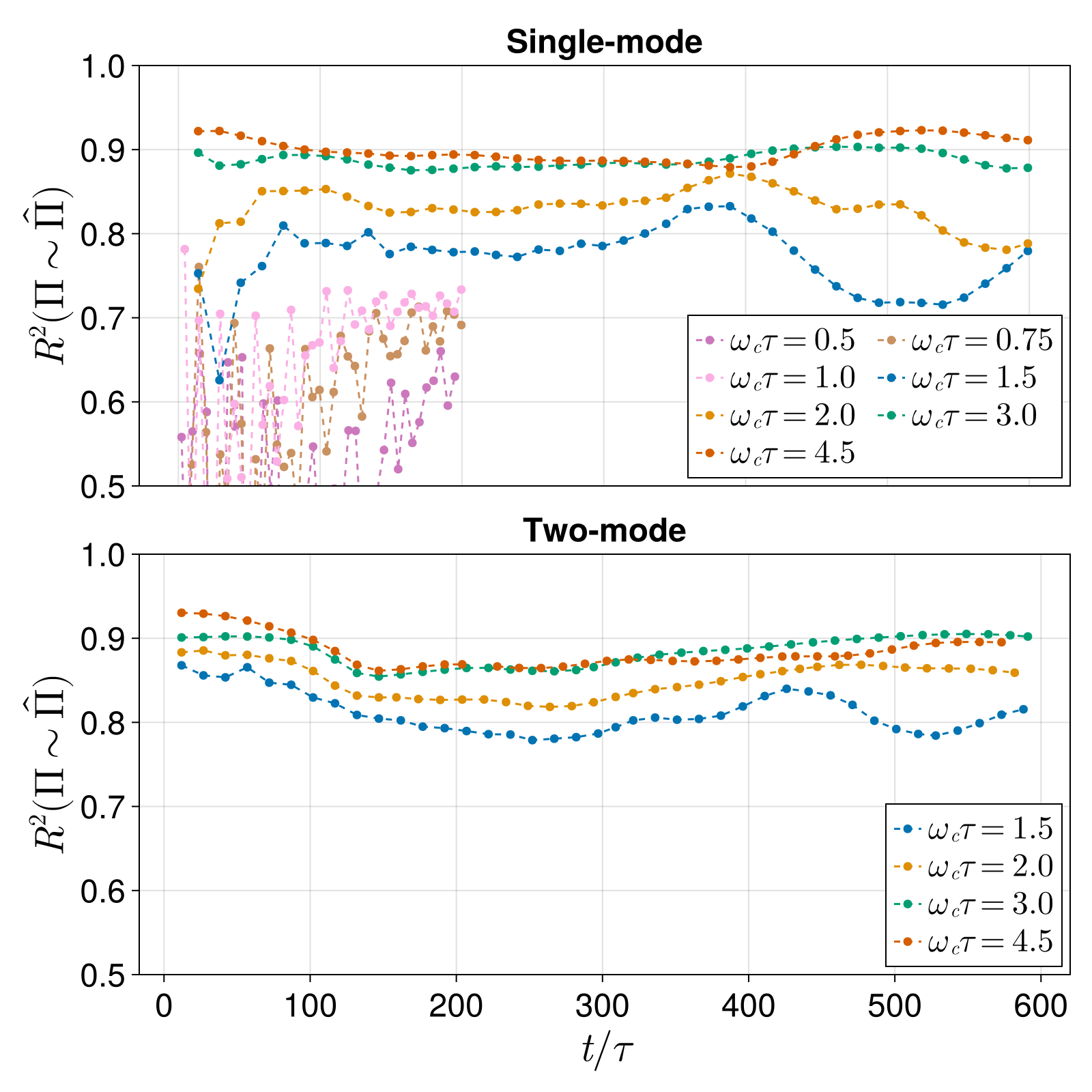}
    \caption{$R^2$ values for \eqref{eqn:Pi_perp} as a function of time for
    simulations A1-A7 (top) and M1-M4 (bottom).
    \label{fig:Pi-perp-R2}
}
\end{figure}

The $R^2$ values for \eqref{eqn:Pi_perp} are plotted as a function of time 
for each of the simulations A1-A7 and M1-M4 listed in Table \ref{table:sf-simulation-table}.
The results are shown in Figure \ref{fig:Pi-perp-R2}.
They indicate that the transport closure improves significantly as $\omega_c \tau$ increases
from 0.5 to 4.5.
The similarity of the $R^2$ traces for $\omega_c\tau = 3.0$ and $\omega_c \tau = 4.5$,
however, suggests that further convergence to the leading-order transport
theory is beyond the ability of our simulations to discriminate.
Confounding factors may include numerical dissipation during the simulation runtime 
as well as errors introduced by gradient approximation during post-processing.
The early time evolution of all simulations is dominated by noise attributable to waves,
which suggests that initializing the stress tensor and heat flux moments
is not sufficient to eliminate startup noise in fully kinetic simulations.

\begin{figure}
    \centering
    \includegraphics[width=0.5\textwidth]{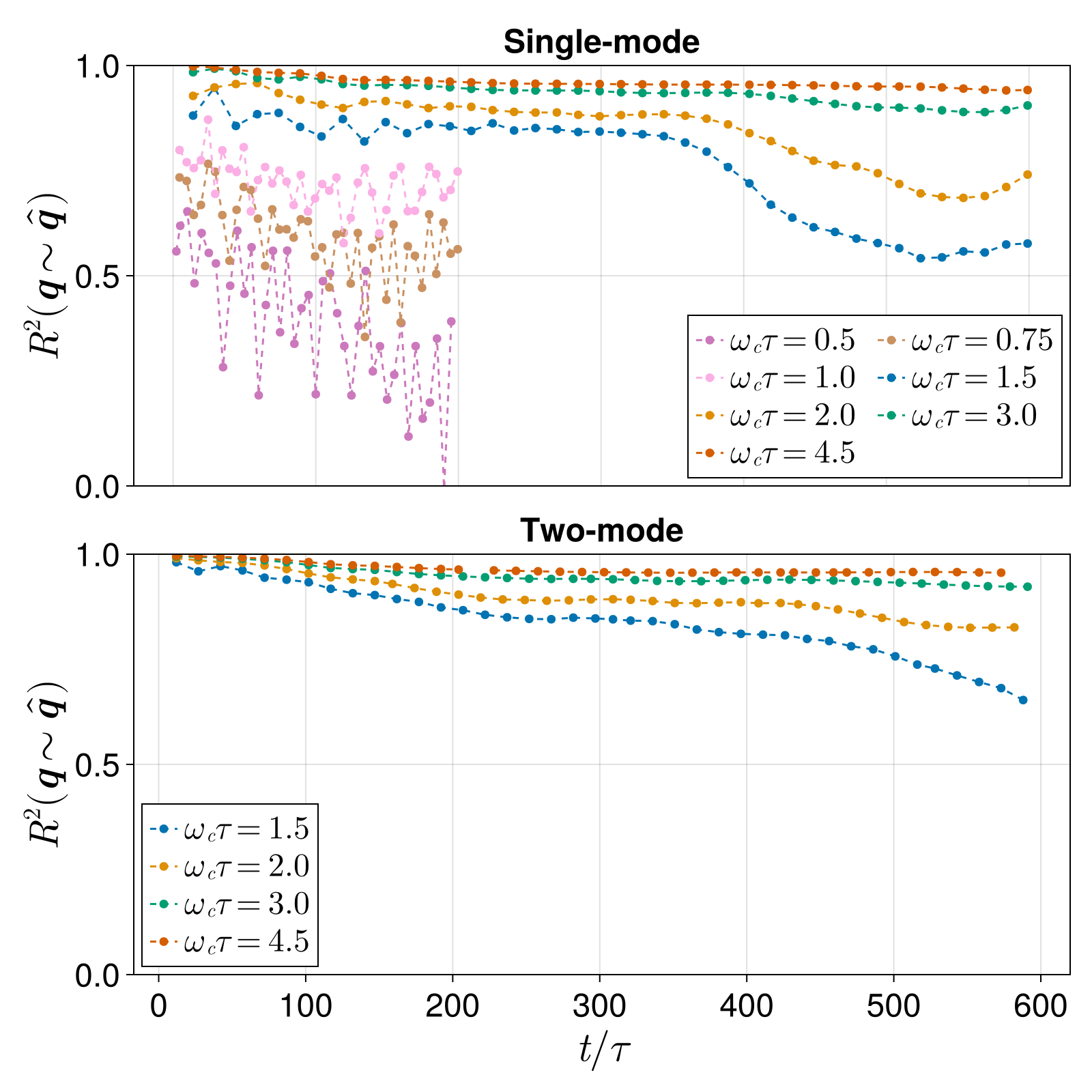}
    \caption{$R^2$ values for \eqref{eqn:qp1-calc} as a function of time for
    simulations A1-A7 (top) and M1-M4 (bottom).
    \label{fig:q-perp-R2}
}
\end{figure}

Figure \ref{fig:q-perp-R2} plots the $R^2$ values for the heat flux closure \eqref{eqn:qp1-calc}.
We observe the same overall pattern of improving agreement as $\omega_c \tau$ increases.
Notably, the overall trend is that the $R^2$ for heat flux is higher than the $R^2$ for
the stress tensor, despite the stress tensor closure being formally of order $\epsilon^2$.
We speculate that this is due to the increased complexity of the stress tensor closure,
and point out that, heuristically, more can ``go wrong'' when using a complex expression to model
the stress tensor compared to the much simpler diamagnetic heat flux closure.

\begin{figure}
    \centering
    \begin{subfigure}[t]{0.49\textwidth}
        \includegraphics[width=\textwidth]{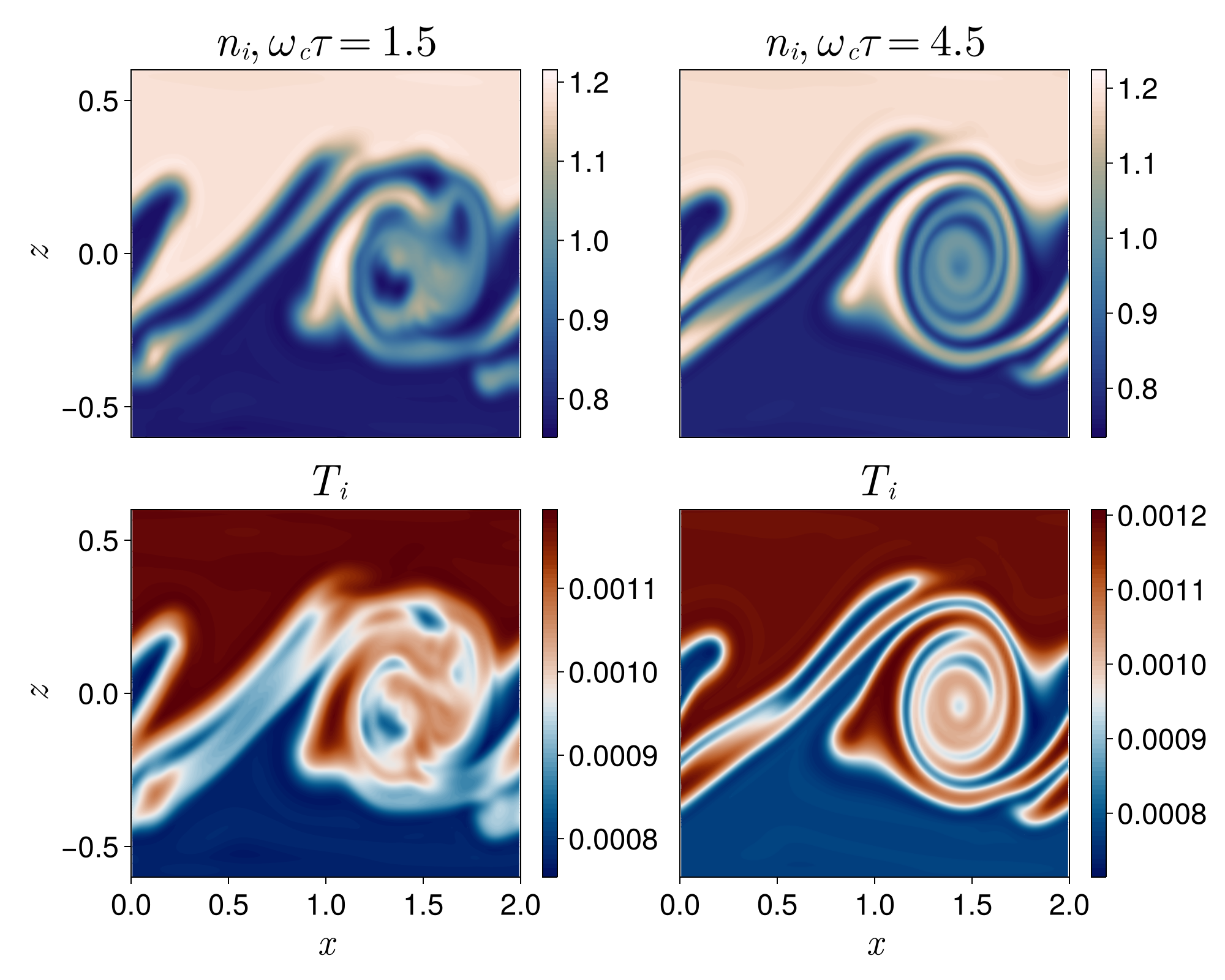}
    \end{subfigure}
    \begin{subfigure}[t]{0.49\textwidth}
        \includegraphics[width=\textwidth]{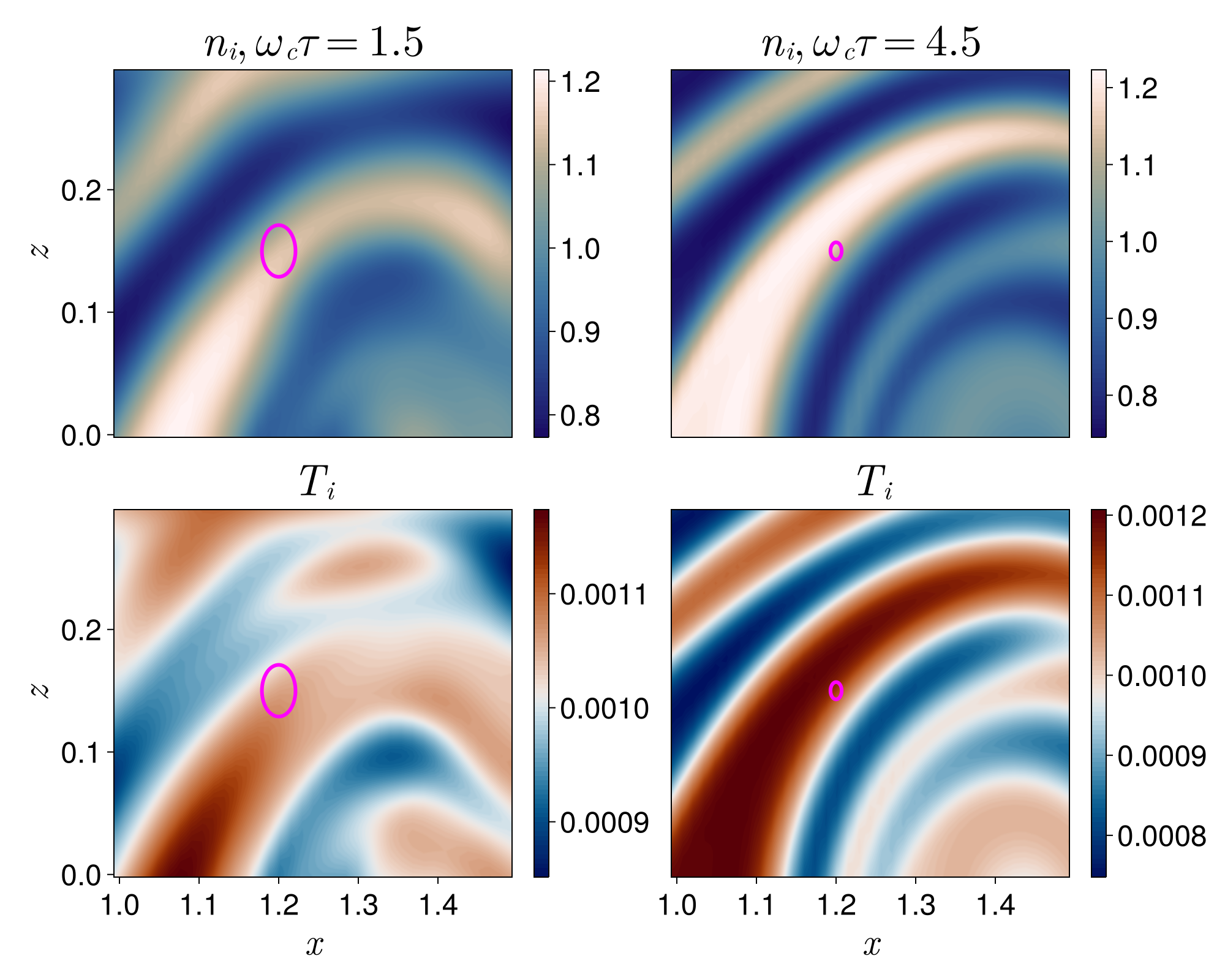}
    \end{subfigure}
    \caption{
        \textbf{Left:}
        Density and temperature contours for cases M1 (first column) and M4 (second column), 
        showing late-time vortex structures and the transition to turbulent flow.
        \textbf{Right:} Zoom of the region $(x, z) \in [1.0, 1.5] \times [0.0, 0.3]$
        with ellipses indicating characteristic ion Larmor radius $r_{Li} = A_i \sqrt{T_{ref}} / (Z_i \omega_c)$.
        \label{fig:two-mode-turbulence}
    }
\end{figure}

The results for single-mode (series A) and two-mode (series M) vorticity are quite comparable.
In general agreement is better for the two-mode series M simulations, which have an $x$ scale
$L_x = 2.0$ of twice that of series A, and thus longer gradient scale lengths in general.
At late times such as $500\tau$ and later, all simulations have become highly distorted and
begun the transition to turbulent mixing.
Figure \ref{fig:two-mode-turbulence} plots the density and 
temperature of cases M1 and M4 at $t = 600\tau$.
The vortex structure is significantly more coherent at this late time for the $\omega_c \tau = 4.5$ case.


\subsection{Affordable adjustment to Braginskii gyroviscous stress}

As described in Section \ref{sec:numerically-feasible-adjustment},
the Braginskii gyroviscous stress closure systematically underestimates
the magnitude of the stress tensor in situations where the pressure gradient
is partially composed of a temperature gradient.
This effect is present whenever there is a temperature gradient in a plasma,
such as in the H-mode \cite{keilhackerHmodeConfinementTokamaks1987} and 
I-mode \cite{manzPhysicalMechanismAccess2020} confinement regimes in tokamaks.
The opposite effect, where the Braginskii gyroviscous stress closure is
a significant overestimation, occurs in the less typical scenario where a temperature
gradient partially or completely balances a density gradient, resulting
in a reduced pressure gradient.
A prototypical example of such a configuration is a magnetized Rayleigh-Taylor unstable
configuration, where a dense plasma is superposed on a hot, less-dense plasma in
an initial balance between thermal pressure forces and some destabilizing force.
Magnetized Rayleigh-Taylor instabilities, and the impact of FLR effects
on their evolution, have been explored in theory and simulation
\cite{hubaRayleighTaylorInstability1998, hubaNonlocalTheoryRayleigh1989, srinivasanRoleElectronInertia2018},
and are relevant to a variety of
applications, including inertial confinement fusion \cite{srinivasanMagneticFieldGeneration2012, srinivasanMitigatingEffectMagnetic2013}.

\begin{figure}
    \centering
    \includegraphics[width=0.8\textwidth]{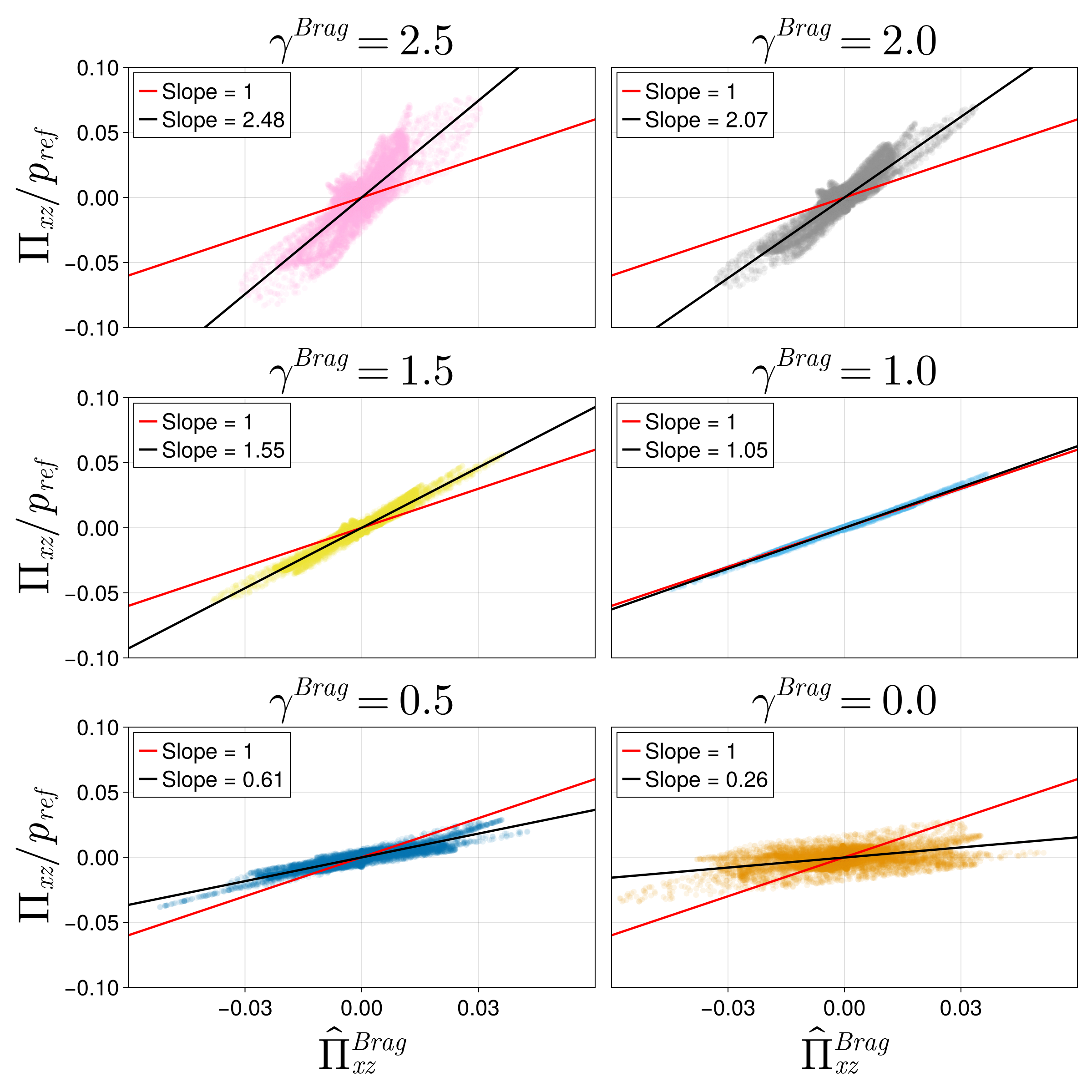}
    \caption{Plots of the Braginskii gyroviscous stress tensor prediction ($x$-axis) versus
    the observed kinetic stress tensor ($y$-axis).
    Black lines are lines of best fit, which have the indicated slopes.
    Best fit slopes show that $\Pi \approx \gamma^{Brag} \hat{\Pi}^{Brag}$ is a decent approximation
    across a range of values of $\gamma^{Brag}$.
    Disagreement is most dramatic in the case $\gamma^{Brag} = 0$, where the bottom right scatterplot shows
    that the Braginskii closure greatly overestimates the magnitude of gyroviscous stress.
    Simulation data are taken from cases C1-C6, which use $\omega_c \tau = 4.0$, at $t = 200\tau$.
    Plotted values are from the $\Pi_{xz}$ component; other components show the same pattern.
    \label{fig:gamma-Brag-Pi_xz}
}
\end{figure}

To examine the role of temperature gradients in setting the gyroviscous
stress, we run simulation series B and C, which vary the value of $\zeta$
from $-0.5$ to $2.0$. This range of $\zeta$ corresponds to values of $\gamma^{Brag}$
from 2.5 to 0.0.
When $\gamma^{Brag} > 1$, the Braginskii gyroviscous stress closure is
expected to be an underestimate.
When $\gamma^{Brag} < 1$, on the other hand, Braginskii overestimates the gyroviscous
stress closure.
A numerical verification of this prediction is shown in Figure \ref{fig:gamma-Brag-Pi_xz},
which plots the Braginskii closure moment $\hat{\Pi}^{Brag}_{xz}$ against the kinetic moment $\Pi_{xz}$
for six different values of $\gamma^{Brag}$ from series C.
In these simulations, shear stress is dominated by shear due to diamagnetic drift, so that
the global factor $\gamma^{Brag}$ is a good estimate of the factor by which the Braginskii
closure over- or underestimates gyroviscous stress.
This is indicated in the slopes of the black lines of best fit, which tend to agree with $\gamma^{Brag}$
in each case.

\begin{figure}
    \centering
    \includegraphics[width=0.8\textwidth]{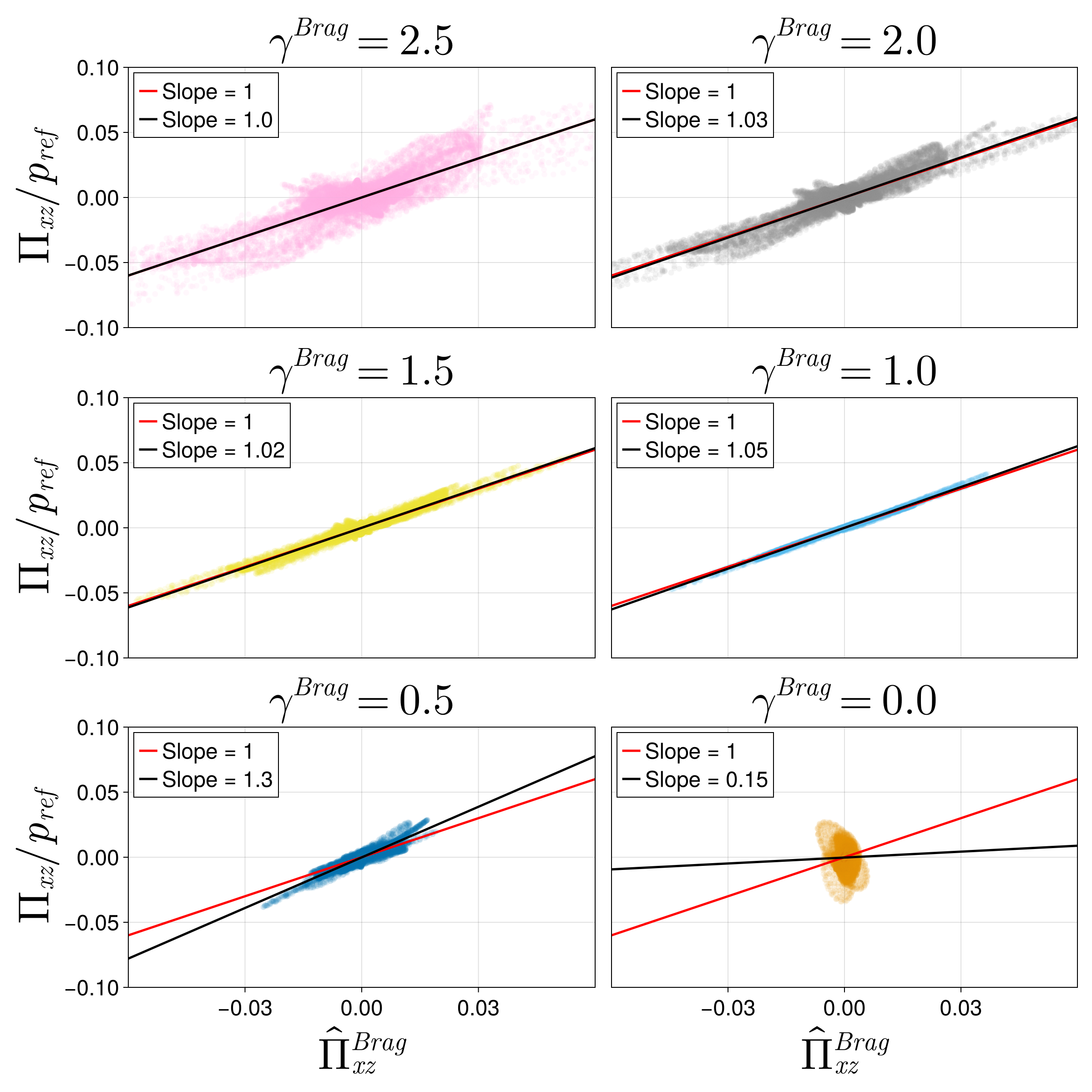}
    \caption{Plots of the adjusted Braginskii gyroviscous stress tensor prediction \eqref{eqn:Pi_adj}, ($x$-axis), versus
    the observed kinetic stress tensor ($y$-axis).
    Black lines are lines of best fit, which have the indicated slopes.
    Compared to Figure \ref{fig:gamma-Brag-Pi_xz}, the quality of the fit is greatly increased for all cases $\gamma^{Brag} \geq 1.0$ (top two rows).
    The adjustment eliminates systematic overestimation of the gyroviscous stress in the case of $\gamma^{Brag} = 0.0$ (bottom right).
    The simulation data are taken from cases C1-C6, which use $\omega_c \tau = 4.0$, at $t = 200\tau$.
    Plotted values are from the $\Pi_{xz}$ component; other components show the same pattern.
    \label{fig:gamma-Brag-Pi_Adj_xz}
}
\end{figure}

The effects of the affordable adjustment to the Braginskii gyroviscous stress, given in
\eqref{eqn:Pi_adj}, are plotted in Figure \ref{fig:gamma-Brag-Pi_Adj_xz}.
The simple adjustment is observed to greatly improve the agreement between the predicted magnitude
of (the $xz$ component of) gyroviscous stress and the observed magnitude.
In particular, the $\gamma^{Brag} = 0.0$ case, which according to Figure \ref{fig:gamma-Brag-Pi_xz} 
is greatly overestimated by the Braginskii closure, is no longer systematically overpredicted by the
adjusted closure.
The cases of $\gamma^{Brag} = 1.5$ and $\gamma^{Brag} = 2.0$ also demonstrate marked improvement
and are quite well predicted by the adjusted closure, whereas the Braginskii closure
underestimates them by $50\%$ and $100\%$, respectively.


\begin{figure}
    \centering
    \includegraphics[width=0.8\textwidth]{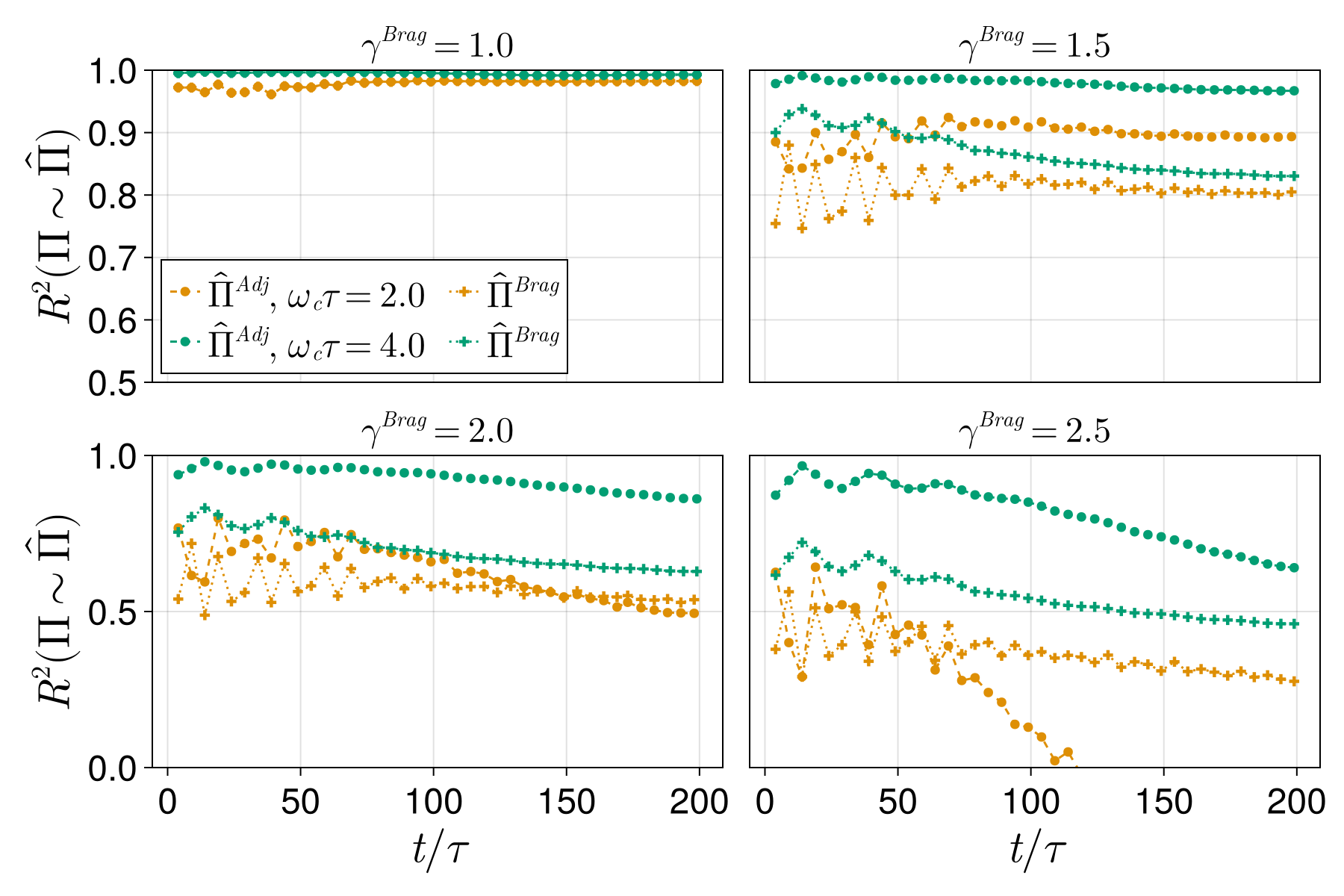}
\caption{
    Comparison of $R^2$ values for unmodified \eqref{eqn:Pi_Brag} (dotted lines) and adjusted \eqref{eqn:Pi_adj} (dashed lines)
    gyroviscous stress tensor closures.
    Plotted simulations are cases B1-B4 ($\omega_c \tau = 2.0$) and C1-C4 ($\omega_c \tau = 4.0$).
    The late-time results show improvement of the adjusted closure $\hat{\Pi}^{Adj}_\perp$ 
    over the unmodified Braginskii closure for the high-magnetization case $\omega_c \tau = 4.0$.
    Agreement of $\hat{\Pi}^{Adj}$ is best for $\gamma^{Brag} = 1.5$ which corresponds to $\zeta = 0.5$,
    i.e. equal density and temperature gradient scale lengths.
    \label{fig:Pi-Adj-R2s}}
\end{figure}

The improvement of the adjusted gyroviscous stress \eqref{eqn:Pi_adj} over the
Braginskii gyroviscous stress closure \eqref{eqn:Pi_Brag} is also reflected in the 
$R^2$ value for $\gamma^{Brag} > 1.0$.
Figure \ref{fig:Pi-Adj-R2s} compares the $R^2$ value of the Braginskii and
adjusted gyroviscous stress closures for simulations B1-B4 and C1-C4.
Improved $R^2$ values are more reliable for the simulations with $\omega_c \tau = 4.0$,
for which leading-order closures are more accurate.
The adjusted closure is less predictive in the case of $\gamma^{Brag} = 2.5$,
which includes the strongest temperature gradient of the cases we consider here.

\begin{figure}
    \centering
    \includegraphics[width=0.8\textwidth]{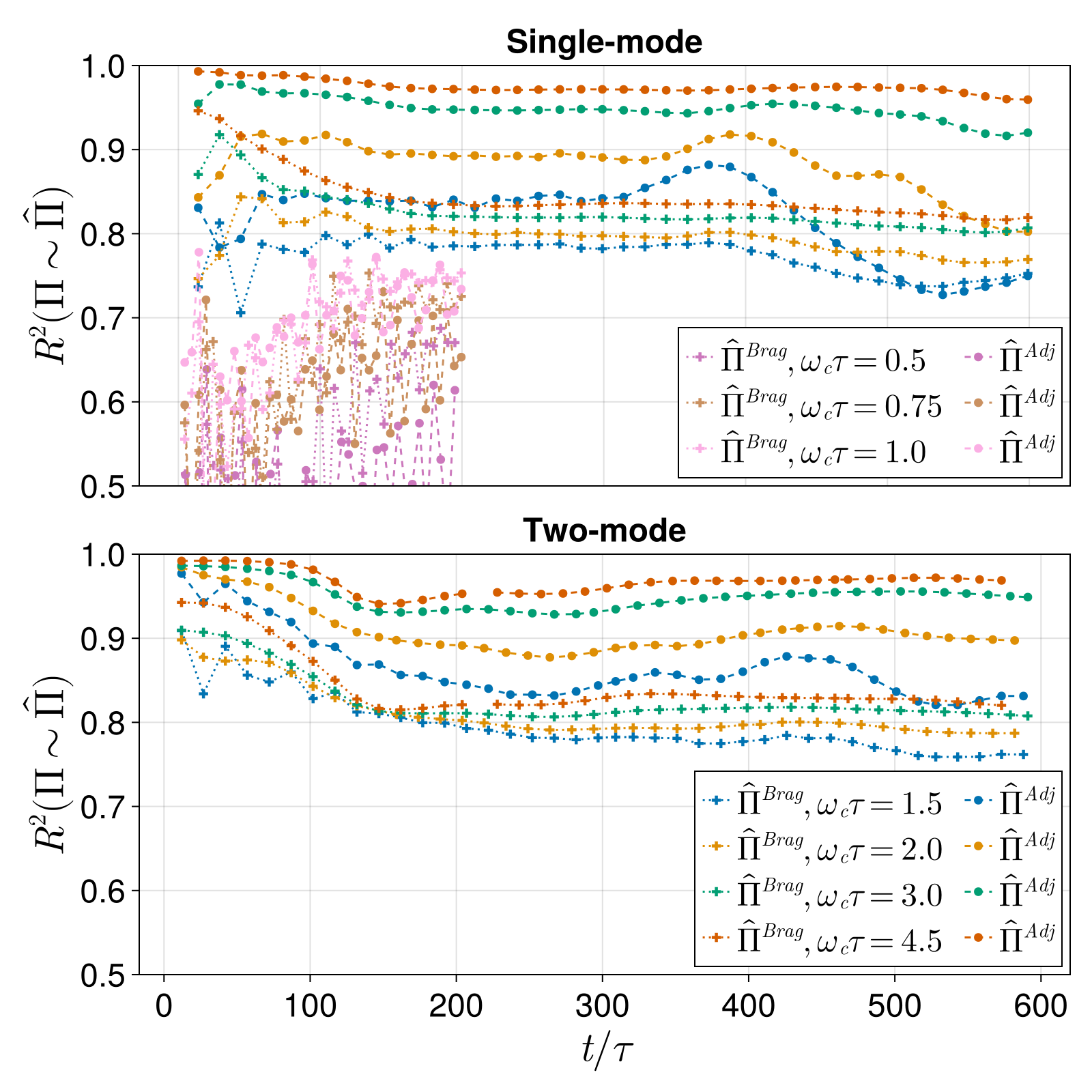}
    \caption{
        Comparison of $R^2$ values for unmodified Braginskii \eqref{eqn:Pi_Brag} (dotted lines)
        and adjusted \eqref{eqn:Pi_adj} (dashed lines)
        gyroviscous stress tensor closures.
        Plotted simulations are cases A1-A7 (top) and M1-M4 (bottom), all of which have $\zeta = 0.5, \gamma^{Brag} = 1.5$.
        \label{fig:onetwo-Pi-Adj-R2s}
    }
\end{figure}

The long-time performance of the adjustment is shown in Figure \ref{fig:onetwo-Pi-Adj-R2s},
which plots $R^2$ values for the Braginskii and adjusted gyroviscous
stress closures for simulations $A1-A7$ and $M1-M4$.
Both simulation series have $\gamma^{Brag} = 1.5$,
corresponding to equal density and temperature gradient scale lengths.
We again observe larger and more consistent improvement for higher magnetization.

\subsection{Higher-order corrections in $\epsilon$}
\label{sec:higher-order}

The transport closures we have examined so far have been only leading-order closures in the small
parameter $\epsilon$, which can be characterized as the ratio of the ion Larmor radius to gradient scale
lengths.
A complete account of kinetic effects, however, naturally requires terms of order $\epsilon^2$ and higher.
We expect that such terms are significant when $\epsilon$ is insufficiently small, resulting
in deviation of the kinetic heat flux and stress tensor from the leading-order closures.
This deviation manifests as reduced $R^2$ for the corresponding closure, as can be seen in the late-time
portion of Figures \ref{fig:Pi-perp-R2}, \ref{fig:q-perp-R2}, and \ref{fig:onetwo-Pi-Adj-R2s},
as well as in the relatively poor agreement of the closure models for low magnetization.

In the collisionless, magnetized limit, higher-order corrections to the heat flux and stress tensor
are of particular interest because the leading-order closure moments do not contribute to diffusion
of heat and dissipative viscous heating, respectively.
For the diamagnetic heat flux, it is simple to see that
\begin{align*}
\hat{\bm{q}} \cdot \nabla T = p \frac{\bm{B} \times \nabla T}{2\Omega_c} \cdot \nabla T = 0,
\end{align*}
so the diamagnetic heat flux does not transport heat along the temperature gradient.
The gyroviscous stress has a similar property, which is revealed by the non-conservative
form of the temperature equation:
\begin{align}
    \label{eqn:sf-temp_eqn_nonconservative}
    \dtt T + \frac{\gamma - 1}{n} (\mathbb{P} : \nabla \bm{u} + \nabla \cdot \bm{q}) = 0.
\end{align}
A simple calculation shows that
\begin{align*}
    \mathbb{W}_3[\bm{u}] : \nabla \bm{u} = 0,
\end{align*}
which means that the Braginskii gyroviscous stress tensor $\hat{\Pi}^{Brag}_\perp$ does not
contribute to dissipative viscous heating.

\begin{figure}
    \centering
    \includegraphics[width=0.8\textwidth]{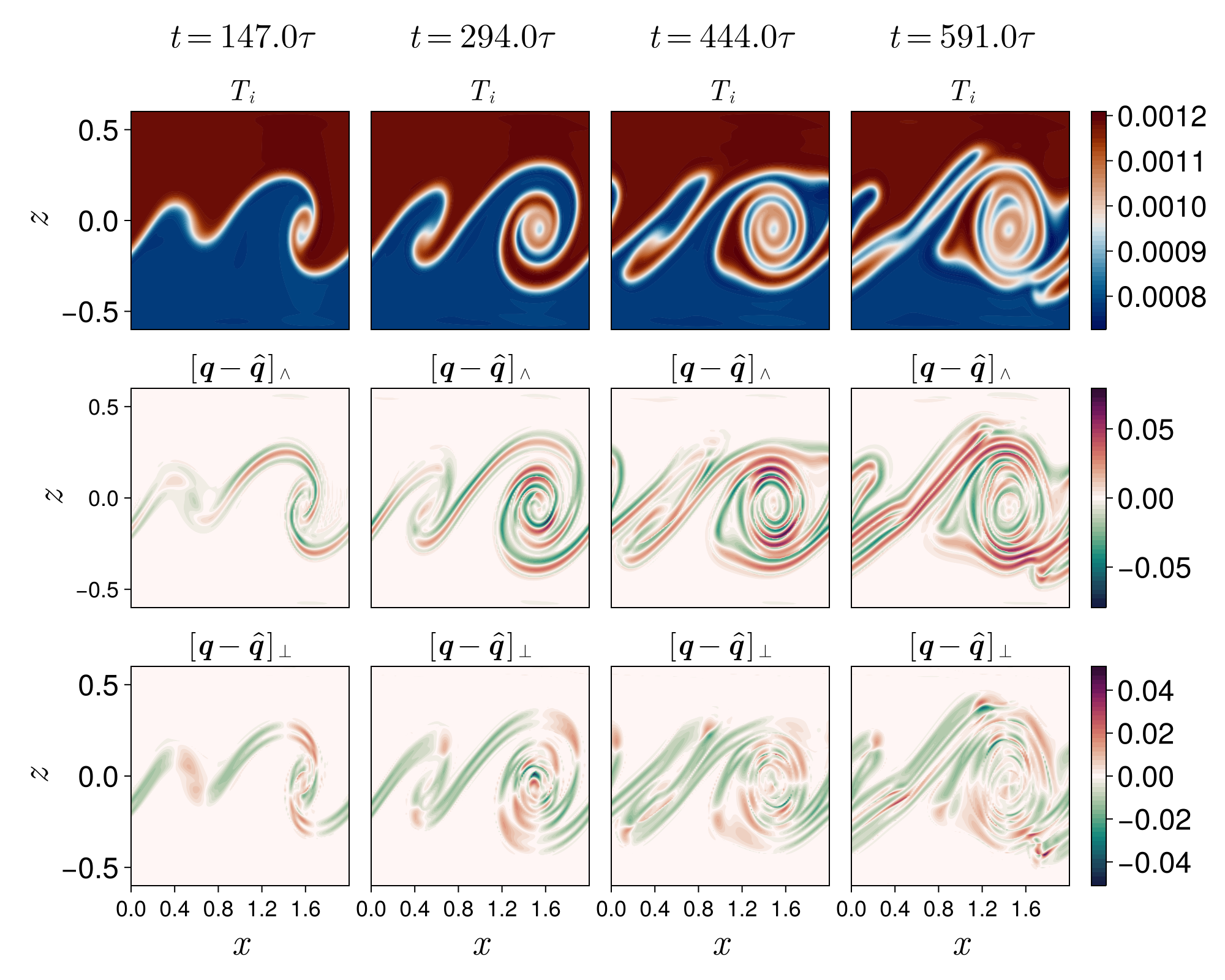}
    \caption{
        Contour plots of the normalized components of the residual $\bm{q} - \hat{\bm{q}}$ for case M3.
        \textbf{First row}: contours of ion temperature.
        \textbf{Second row}: contours of $[\bm{q} - \hat{\bm{q}}]_\wedge$, defined as the component of $\bm{q} - \hat{\bm{q}}$
        in the direction of $\bm{B} \times \nabla T$.
        Contours plotted in units of the reference free-streaming heat flux $\bm{q}_{fs} = v_{ti} p_{ref}$.
        Spatial structure in the shear layer, in particular multiple sign changes crossing the shear
        layer, indicate that second- or third-order derivative polynomials of temperature likely
        play a role in the leading-order residual. This is consistent with the structure of the
        asymptotic expansions at higher order in $\epsilon$.
        \textbf{Third row}: contours of $[\bm{q} - \hat{\bm{q}}]_\perp$, defined as the component of $\bm{q}$
        in the direction of $\nabla T$. 
        Contours plotted in units of $\bm{q}_{fs} = v_{ti} p_{ref}$
        Negative values of this component indicate diffusion of heat
        while positive values indicate anti-diffusion of heat.
        The sign of diffusion exhibits a clear dependence on the slope of the vortex in the $x-z$ plane,
        consistent with ion inertial effects (see discussion). \label{fig:q_residuals_plot1}}
\end{figure}

To better understand the role of higher-order corrections to the closure moments, we plot
the residuals of the leading-order closures.
The heat flux at each point $\bm{x}$ can be decomposed into a diamagnetic component orthogonal to $\nabla T$ and a perpendicular component which is parallel to $\nabla T$.
We define the normalized component decomposition of the residual $\bm{q} - \hat{\bm{q}}$ in the following way:
\begin{align}
    [\bm{q} - \hat{\bm{q}}]_\wedge = \frac{(\bm{q} - \hat{\bm{q}}) \cdot (\bm{B} \times \nabla T)}{A_i |B||\nabla T| v_{ti}^{3/2}}, \quad
    [\bm{q} - \hat{\bm{q}}]_\perp = \frac{(\bm{q} - \hat{\bm{q}}) \cdot \nabla T}{A_i |\nabla T| v_{ti}^{3/2}}.
\end{align}
These expressions are normalized by the free-streaming heat flux limit, which is $A_i v_{ti}^{3/2}$.
Figure \ref{fig:q_residuals_plot1} plots these expressions along with the ion temperature
for case M3 at four different times.
The residual plots reveal coherent structure well into the nonlinear phase.
Comparing the residuals with plots of temperature indicate that the heat flux
closure residual aligns with regions of high curvature of $T_i$, consistent
with the residual being well-described by a second or third-degree derivative
polynomial in $T_i$.
Such expressions arise at higher order in the asymptotic expansion procedure.
However, the size of such higher-order corrections naturally has a quadratic
or cubic dependence on the inverse temperature gradient scale length.
Turbulent flow, being characterized by high-wavenumber spatial features in density
and temperature, therefore cannot be expected to conform to the leading-order closure expressions.
Moreover, the presence of spatially localized features in the residual plots highlights
the importance of using local estimates for closure applicability, rather than global correction factors
based on a single problem parameter.

\begin{figure}
    \centering
    \includegraphics[width=0.8\textwidth]{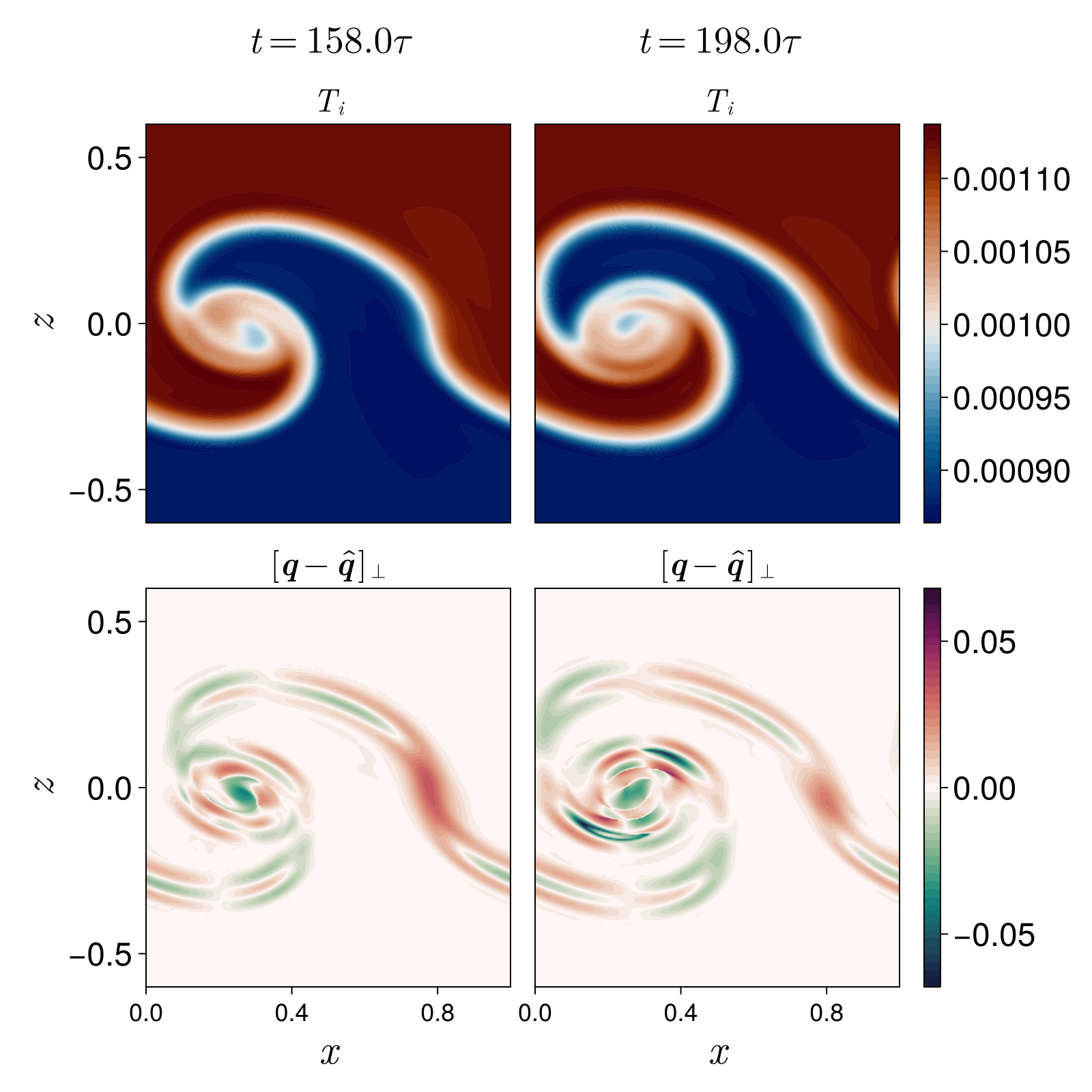}
    \caption{
        Contour plots of the normalized perpendicular component of the residual $\bm{q} - \hat{\bm{q}}$ 
        for case S1. 
        \textbf{First row}: contours of ion temperature.
        \textbf{Second row}: contours of $[\bm{q} - \hat{\bm{q}}]_\perp$, defined as the component of $\bm{q} - \hat{\bm{q}}$
        in the direction of $\nabla T$.
        Contours are plotted in units of the reference free-streaming heat flux $\bm{q}_{fs} = v_{ti} p_{ref}$.
        Dependence of sign of diffusion on the slope of the vortex matches
        the dependence observed in simulation M3
        (compare to the third row of Figure \ref{fig:q_residuals_plot1}.)
\label{fig:q_residuals_reversed_shear}}
\end{figure}

Examining the third row of Figure \ref{fig:q_residuals_plot1} in more detail, we note that the sign of
the perpendicular (along-gradient) heat flux residual has a clear dependence on the slope of 
the rollup in the $x-z$ plane.
Case S1, which has a reversed shear direction, is plotted in Figure \ref{fig:q_residuals_reversed_shear}
and shows the same trend.
A negative sign of $[\bm{q} - \hat{\bm{q}}]_\perp$ indicates a heat flux vector in the 
opposite direction of $\nabla T$, and thus perpendicular diffusion of heat.
On the other hand, a positive sign of $[\bm{q} - \hat{\bm{q}}]_\perp$ indicates
anti-diffusion of heat.

We hypothesize that this is attributable to higher-order corrections to heat flux
associated with ion inertia: as the slope of the vortex increases, the heat flux vector, which
is initially in the negative $x$ direction, lags behind the changing $\bm{B} \times \nabla T$ direction
and acquires nonzero components directed parallel to $\nabla T$. 
To see the origin of this effect, we write the perpendicular moment equation for the 
heat flux in non-conservative form with index notation:
\begin{align}
    \label{eqn:sf-qeqn-index}
    \frac{\mathrm{d}}{\mathrm{d} t} q_i + \partial_j H_{ij} + \partial_j u_j q_i = \frac{Z}{A} \left[ (\omega_p \tau E_j + \omega_c \tau \epsilon_{jmn} u_m B_n) \left[ \delta_{ij} \pp + \mathbb{P}_{ij} \right] + \omega_c \tau \epsilon_{imn} q_m B_n + \omega_c \tau \epsilon_{jmn} B_n Q_{ijm} \right],
\end{align}
where
\begin{align*}
    H_{ij} = \frac{A}{2} \left\langle w_i w_j w_k w_k f \right\rangle_v, \quad Q_{ijm} = \left\langle w_i w_j w_m f \right\rangle_v.
\end{align*}
Here we are using the notation $\bm{w} = \vp - \up$ and $\left\langle \cdot \right\rangle_v = \int \cdot \dvp$.
Substituting the Maxwellian moments $H_{ij} = \frac{2pT}{A} \delta_{ij}$, $Q_{ijm} = 0$, and $\mathbb{P}_{ij} = \pp\delta_{ij}$, we can simplify the expression and rewrite in vector notation:
\begin{align}
    \label{eqn:sf-qeqn-vector}
    \dtt \bm{q}_\perp + \frac{2}{A} \nabla (\pp \Tp) + (\nabla \cdot \up) \bm{q}_\perp = \frac{Z}{A} \left[ ( \omega_p \tau \bm{E} + \omega_c \tau \up \times \bm{B}) \cdot \left[ 2\pp \mathbb{I} \right] + \omega_c \tau \bm{q}_\perp \times \bm{B} \right].
\end{align}
Substituting the leading-order drift velocity $\up^1$ into \eqref{eqn:sf-qeqn-vector} and
neglecting flow compressibility, we get an equation for the leading-order heat flux,
\begin{align*}
2\nabla(\pp \Tp) = Z \left( 2\frac{\nabla \pp}{\np} \pp + \omega_c \tau \bm{q}^1_\perp \times \bm{B}\right),
\end{align*}
whose solution is the diamagnetic heat flux \eqref{eqn:qp1-calc} up to order $\epsilon^2$.
At the subsequent order, we substitute the polarization drift $\up^p$ on the right-hand side
and obtain
\begin{align*}
    \dtt \bm{q}_\perp^1 = \frac{Z}{A} \left[ (\omega_c \tau \up^p \times \bm{B}) (2\pp) + \omega_c \tau \bm{q}_\perp^p \times \bm{B} \right],
\end{align*}
or
\begin{align*}
\bm{q}_\perp^p = \frac{A}{Z \omega_c \tau |B|^2} \left( \dtt \bm{q}_\perp^1 \right) \times \bm{B} - 2\pp \up^p.
\end{align*}
The heat flux at next-to-leading order, $\bm{q}_\perp^p$, is therefore seen to be associated with ion inertial effects,
via the time derivative of the leading-order heat flux as well as the ion polarization drift.
Polarization drifts were observed to drive charge accumulation in the Kelvin-Helmholtz instability
\cite{vogmanTwofluidKineticTransport2020}.
The simulations conducted here suggest that similar physics may drive heat accumulation,
via along-gradient heat fluxes, in Kelvin-Helmholtz-like vortex structures.

For the stress tensor closure, we split the residual $\Pi_\perp - \hat{\Pi}_\perp$ into two
components, one in the direction of $\mathbb{W}_1[\bm{u}]$ and the other in the direction of
$\mathbb{W}_3[\bm{u}]$.
Note that $\mathbb{W}_1[\bm{u}]$ is defined as
\begin{align*}
    \mathbb{W}_1[\bm{u}] = \begin{pmatrix}
        \partial_x u_x - \partial_z u_z & \partial_x u_z + \partial_z u_x \\
        \partial_z u_x + \partial_x u_z & \partial_z u_z - \partial_x u_x
    \end{pmatrix},
\end{align*}
and satisfies $\mathbb{W}_1 : \mathbb{W}_3 = 0$.
Since $\Pi$ is a symmetric, trace-free tensor, it has two degrees of freedom and is therefore
uniquely determined by its magnitude in the direction of $\mathbb{W}_1$ and $\mathbb{W}_3$, respectively.
By the same token, $\Pi : \mathbb{W}_1$ indicates the proportion of stress that contributes to
dissipative heating via \eqref{eqn:sf-temp_eqn_nonconservative}, while $\Pi : \mathbb{W}_3$
indicates the proportion of perpendicular stress that contributes to transverse but non-dissipative
transport of momentum.

\begin{figure}
    \centering
    \includegraphics[width=1.0\textwidth]{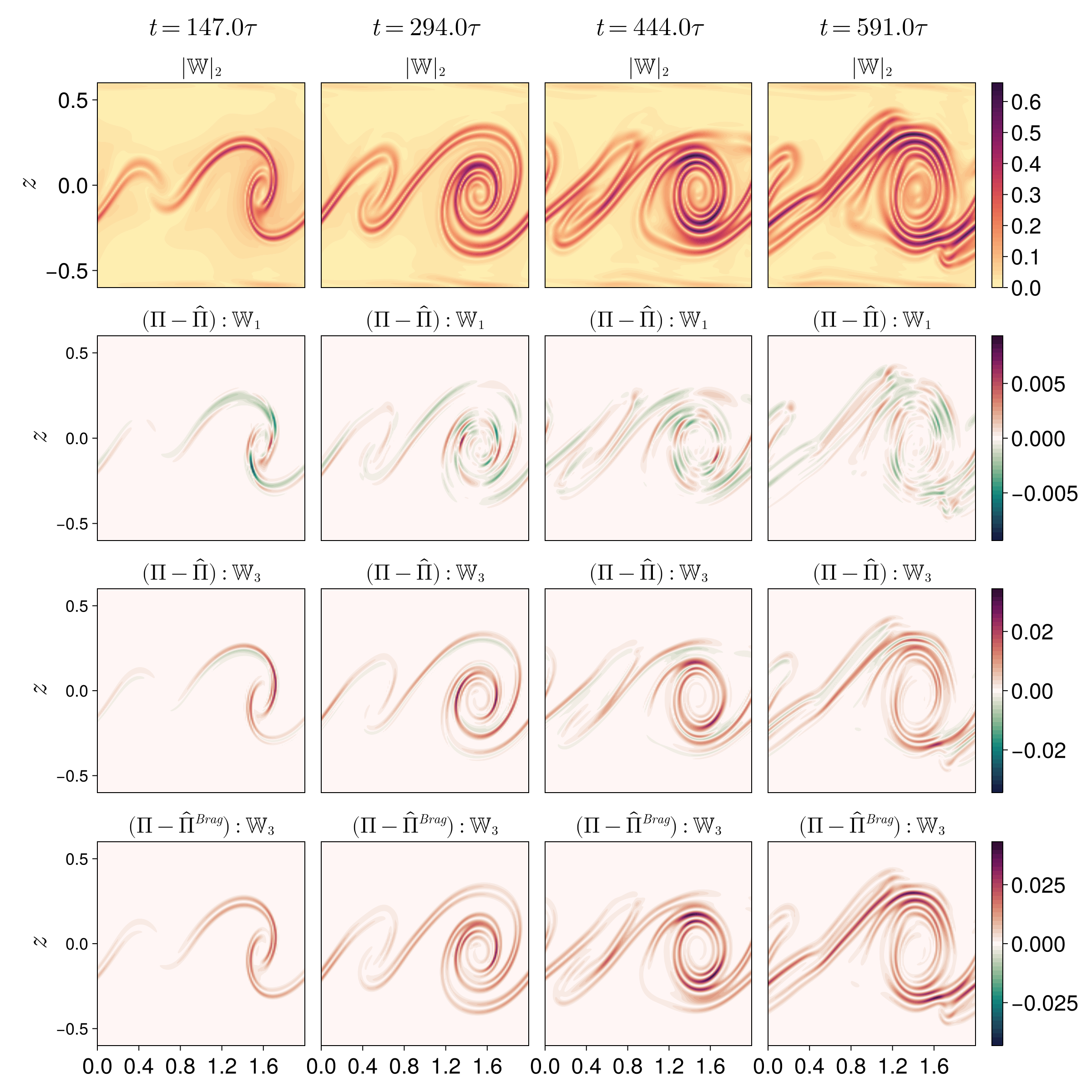}
    \caption{
            Components of the stress tensor residual from simulation M3, with $\omega_c \tau = 3.0$.
            \textbf{First row:} $L^2$ norm of $\nabla \bm{u}$ in units of $v_{ti} / \alpha$.
            \textbf{Second row:} the magnitude of the component of $\Pi - \hat{\Pi}$ in the direction of $\mathbb{W}_1[\bm{u}]$.
            \textbf{Third row:} magnitude of the component of $\Pi - \hat{\Pi}$ in the direction of $\mathbb{W}_3[\bm{u}]$.
            \textbf{Fourth row:} magnitude of the Braginskii gyroviscous stress residual $\Pi - \hat{\Pi}^{Brag}$ in
            the direction of $\mathbb{W}_3[\bm{u}]$.
            Note that colorbars in rows 2-4 are centered at 0 to facilitate interpretation of signed quantities.
            Rows 2 and 3 show no bias in the direction of $\mathbb{W}_1$, but a significant bias in the direction
            of $\mathbb{W}_3$, indicating that gyroviscous transport of transverse momentum is greater in kinetic
            solutions than predicted by leading-order closures.
            Row 4 illustrates the consistent underestimation of kinetic stress by the Braginskii closure.
\label{fig:Pi-residuals-plot1}}
\end{figure}

The results of this analysis are plotted in Figure \ref{fig:Pi-residuals-plot1}.
The first row plots contours of the norm of the shear stress tensor, $|\mathbb{W}|$, in units of a reference 
shear frequency which we define as $\omega_s = v_{ti} / \alpha$, the ratio of the thermal velocity to initial 
interface width.
As the vortex evolves, the magnitude and complexity of the velocity shear structures increases, presenting
increased difficulty for leading-order gyroviscous stress closures.

The second row of Figure \ref{fig:Pi-residuals-plot1} plots $(\Pi : \mathbb{W}_1[\bm{u}]) / (p_{ref} \omega_s)$.
Notably, this quantity exhibits no discernable bias in one direction or another and is quite small,
remaining less than $1\%$ for the entire simulation lifetime.
This indicates that systematic errors in the stress tensor closure for this simulation do not
omit substantial amounts of dissipative viscous stress.
The third row of the figure indicates the opposite conclusion for the component of $\Pi$ in the 
direction of $\mathbb{W}_3[\bm{u}]$, which does exhibit a persistent bias in the positive direction.
This indicates that the leading-order gyroviscous stress closure systematically underestimates the stress, despite
the inclusion of the heat flux correction term.
The final row of Figure \ref{fig:Pi-residuals-plot1} plots the same quantity for the Braginskii gyroviscous
stress.
Comparing the third and fourth rows, we conclude that while the leading-order gyroviscous stress is an
improvement over Braginskii and substantially reduces the underestimation error, it does not eliminate it.

\section{Conclusion}
\label{sec:conclusion}

The drift ordering limit of the Vlasov equation is rigorously defined via a consistent
non-dimensionalization.
The resulting scaling is applicable to a variety of highly magnetized, low-beta plasmas
such as tokamaks \cite{hazeltineFourfieldModelTokamak1985}
and those that form around magnetically insulated transmission lines 
\cite{vogmanTwofluidKineticTransport2020, spielmanDesignMagneticallyInsulated2019, bakshaevStudyDynamicsElectrode2007}.
Most importantly, the scaling considered encompasses plasmas with collision frequencies
that are arbitrarily small.
A semi-fluid theory for such plasmas is derived by taking moments of $f$ with respect
to powers of the perpendicular velocity while leaving the parallel velocity dependence kinetic.

Based on the assumption of a leading-order distribution function which is gyrotropic and 
Maxwellian in $\vp$, the Vlasov equation in the drift ordering is expanded in powers of $\epsilon$.
The expansion depends on Fredholm solvability conditions which are naturally satisfied by
physically plausible collision operators such as the Landau operator.
In order to retain a time-dependent momentum equation despite the drift velocity being an order $\epsilon$
quantity, the kinetic equation is manipulated to locate all perpendicular momentum in the
first-order distribution function correction.
By expanding to order $\epsilon^2$, the leading-order perpendicular heat flux and stress tensor
closures are determined.
Heat flux is found to be diamagnetic and non-diffusive. 
The leading-order stress is
found to be composed of the classical gyroviscous stress plus a correction due to the order-$\epsilon$ 
distortion of $f$ in the presence of temperature gradients.
The correction indicates that the classical gyroviscous stress closure is an underestimate
in situations where the pressure gradient is partially composed of a temperature gradient.
To enable MHD and multi-fluid simulation codes
\cite{breslauPropertiesM3DC1Form2009, shumlakAdvancedPhysicsCalculations2011, giacominGBSCodeSelfconsistent2022, sovinecNonlinearMagnetohydrodynamicsSimulation2004, fryxellFLASHAdaptiveMesh2000}
to easily account for this correction, a numerically affordable
adjustment to the Braginskii gyroviscous stress is proposed based on an estimate of the factor
\begin{align*}
    \gamma^{Brag} = \frac{|\nabla p / p + \nabla T / T|}{|\nabla p / p|}.
\end{align*}

To explore the quantitative importance of the disagreement between the Braginskii and drift ordering
stress tensor closures, an electrostatic Vlasov simulation code is developed for
straight-line magnetic fields and slab geometries with one direction of non-periodicity.
The code uses a spectral representation of velocity space based on Hermite polynomials and
a Fourier pseudospectral collocation representation of the periodic dimensions of physical space.
The non-periodic $z$ dimension is represented using a high-order finite difference discretization.
The Vlasov solver is applied to a family of magnetized initial conditions which exhibit
sheared flow driven by $E \times B$ drifts, vorticity, and density and
temperature gradients.
To facilitate exploration of the key parameters governing accuracy of the transport closures
and the importance of the disagreement between the Braginskii and drift ordering stress
tensor closures, the initial conditions are parameterized by magnetization and by
the relative size of the density and temperature gradient length scales.

Simulation results show that the drift ordering closure exhibits convergence
to the observed kinetic moments with increasing magnetization.
For low magnetizations (characterized by the ratio of cyclotron to plasma frequency), 
the closures are found to be unreliable
and to leave much of the variation in the kinetic moments unexplained.
For magnetizations of $\omega_c \geq 1.5\omega_p$, the closures are predictive and explain the
majority of the spatial variation in the kinetic moments.
The under- and over-estimation committed by the classical Braginskii gyroviscous stress closure
in the presence of temperature gradients is evaluated for a range of values of $\gamma^{Brag}$.
Descriptive statistics validate the prediction of the drift ordering transport theory regarding
the direction and approximate magnitude of the Braginskii closure's error.
The affordable adjustment to Braginskii corrects these errors for the highly magnetized case
of $\omega_c = 4.0 \omega_p$, and is slightly
less effective at reducing unexplained variance for the moderately magnetized $\omega_c = 2.0 \omega_p$.

Residuals of the transport closures have complex spatial structure indicating
the importance of higher-order contributions to the heat flux and stress tensor.
Analysis of the components of heat flux residuals parallel to the temperature gradient
indicates that second-order ion inertial physics may play a role in diffusive and anti-diffusive transport
of heat along the vortex roll-up.
Diamagnetic heat flux residuals perpendicular to both $\bm{B}$ and $\nabla T$ exhibit
spatial structure consistent with second- and third-order derivative polynomials of
temperature, as would arise in higher-order asymptotic expansions in $\epsilon$.

The residuals of $\Pi$ are analyzed in terms of their components in the direction of the viscous shear stress
$\mathbb{W}_1$ and the gyroviscous shear stress $\mathbb{W}_3$.
It is found that the kinetic shear stress is almost entirely gyroviscous and therefore
non-diffusive.
Moreover, the higher-order contributions summarized in the residual demonstrate significant
bias in the direction of $\mathbb{W}_3$, indicating that both the drift ordering and Braginskii
closures commit systematic underestimation of the gyroviscous stress when $|\nabla p \cdot \nabla T| > 0$,
although the drift ordering closure is a major improvement compared to the Braginskii closure.

Future work in this direction should include the numerical evaluation of the gyroviscous stress
closures we present here in the fast dynamics regime, in which drift velocities are a
substantial fraction of the thermal velocity: $|\bm{u}_s| \gtrsim 0.5 v_{ti}$.
The Kelvin-Helmholtz simulations performed in Refs. \onlinecite{vogmanTwofluidKineticTransport2020} and 
\onlinecite{vogmanHighfidelityKineticModeling2021}, for example, include drift velocities of approximately
this magnitude driven by the combination of $E \times B$ and diamagnetic drifts.
Despite being initialized isothermally, these Kelvin-Helmholtz simulations exhibit significant
temperature gradients in the nonlinear phase due to non-adiabatic effects.
It is observed that the Braginskii gyroviscous stress closure is not a perfect approximation of
the kinetic stress for these simulations, particularly in the nonlinear regime \cite{vogmanHighfidelityKineticModeling2021}.
The importance of such dynamically generated temperature gradients in setting the magnitude of gyroviscous stress remains
unclear, given that the drift velocities in these simulations are large enough to call into
question the appropriateness of a gyrotropic Maxwellian ansatz for $f^0$.
A study similar to the one conducted here, which varies the shear velocity parameter $\bm{u}_s$,
would shed further light on the regime of validity of the drift ordering closure and its numerically affordable
approximation, the adjusted Braginskii closure.

\begin{acknowledgments}
The information, data, or work presented herein is based upon work supported by the National Science Foundation under Grant No. PHY-2108419. JH's research was also partially supported by AFOSR grant FA9550-21-1-0358, NSF grant DMS-2409858, and DOE grant DE-SC0023164.
\end{acknowledgments}

\section*{Author Declarations}
The authors have no conflicts to disclose.

\section*{Data Availability Statement}
The data that support the findings of this study are available from the corresponding author upon reasonable request.

\bibliography{../LowBetaAsymptotics}

\appendix

\section{Definition of shear stress tensor components}

This section reproduces the definitions of $\mathbb{W}_0$ through $\mathbb{W}_4$ from Ref. \onlinecite{braginskiiTransportProcessesPlasma1965}.

The rate-of-strain tensor $\mathbb{W}$ is defined in terms of the velocity field $\bm{u}$ as
\begin{align*}
\mathbb{W} = \nabla \bm{u} + (\nabla \bm{u})^T - \frac{2}{3} (\nabla \cdot \bm{u}) \mathbb{I},
\end{align*}
where $\mathbb{I}$ is the $3 \times 3$ identity tensor.

In index notation, the tensors $\mathbb{W}_{0\dots 4}$ are defined as follows:
\begin{align*}
    [\mathbb{W}_0]_{ij} &= \frac{3}{2} \left( b_i b_j - \frac{1}{3} \delta_{ij} \right) \left( b_k b_l - \frac{1}{3} \delta_{kl} \right)  \mathbb{W}_{kl}, \\
    [\mathbb{W}_1]_{ij} &= \left( \delta^\perp_{ik} \delta^\perp_{jl} + \frac{1}{2} \delta^\perp_{ij} b_k b_l \right)  \mathbb{W}_{kl}, \\
    [\mathbb{W}_2]_{ij} &= \left( \delta^\perp_{ik} b_j b_l + \delta^\perp_{jl} b_i b_k \right)  \mathbb{W}_{kl}, \\
    [\mathbb{W}_3]_{ij} &= \frac{1}{2} \left( \delta^\perp_{ik} \epsilon_{jml} + \delta^\perp_{jl} \epsilon_{imk} \right) b_m \mathbb{W}_{kl} \\
    [\mathbb{W}_4]_{ij} &= \left( b_i b_k \epsilon_{jml} + b_j b_l \epsilon_{imk} \right) b_m \mathbb{W}_{kl},
\end{align*}
where $\delta^\perp_{ij} = \delta_{ij} - b_i b_j$ and $\epsilon_{imk}$ is a Levi-Civita symbol.

In a right-handed coordinate triplet $(x, y, \parallel)$, where $x$ and $y$ are coordinates for the perpendicular
directions, we have
\begin{align*}
\bm{b} = (0, 0, 1)^T, \quad \delta^\perp = \begin{pmatrix}
    1 & 0 & 0 \\
    0 & 1 & 0 \\
    0 & 0 & 0
\end{pmatrix},
\end{align*}
and the tensors have the following explicit expressions:
\begin{align*}
\mathbb{W}_0 &= \begin{pmatrix}
    \frac{1}{2} (\mathbb{W}_{xx} + \mathbb{W}_{yy}) & 0 & 0 \\
    0 & \frac{1}{2} (\mathbb{W}_{xx} + \mathbb{W}_{yy}) & 0 \\
    0 & 0 & \mathbb{W}_{\parallel \parallel}
\end{pmatrix} \\
    \mathbb{W}_1 &= \begin{pmatrix}
        \frac{1}{2} (\mathbb{W}_{xx} - \mathbb{W}_{yy}) & \mathbb{W}_{xy} & 0 \\
        \mathbb{W}_{yx} & \frac{1}{2} (\mathbb{W}_{yy} - \mathbb{W}_{xx}) & 0 \\
        0 & 0 & 0
    \end{pmatrix} \\
        \mathbb{W}_2 &= \begin{pmatrix}
            0 & 0 & \mathbb{W}_{x\parallel} \\
            0 & 0 & \mathbb{W}_{y\parallel} \\
            \mathbb{W}_{\parallel x} & \mathbb{W}_{\parallel y} & 0
        \end{pmatrix} \\
            \mathbb{W}_3 &= \begin{pmatrix}
                -\mathbb{W}_{xy} & \frac{1}{2}(\mathbb{W}_{xx} - \mathbb{W}_{yy}) & 0 \\
                \frac{1}{2} (\mathbb{W}_{xx} - \mathbb{W}_{yy}) & \mathbb{W}_{xy} & 0 \\
                0 & 0 & 0
            \end{pmatrix} \\
            \mathbb{W}_4 &= \begin{pmatrix}
                0 & 0 & -\mathbb{W}_{y\parallel} \\
                0 & 0 & \mathbb{W}_{x \parallel} \\
                -\mathbb{W}_{\parallel y} & \mathbb{W}_{\parallel x} & 0
            \end{pmatrix}
\end{align*}

\section{Numerical discretization of the kinetic-fluid hybrid model}
\label{eqn:numerical_methods}

In this section we describe the numerical methods used to solve the ion Vlasov equation \eqref{eqn:ion-vlasov} and the electron ``drift-advection'' equation \eqref{eqn:electron-drift-advection}
in two perpendicular dimensions (``2D2V'').

The kinetic ion species is discretized using a Hermite spectral discretization in velocity space.
Hermite spectral methods have been used for the velocity dimension of the Vlasov equation before
\cite{schumerVlasovSimulationsUsing1998, delzannoMultidimensionalFullyimplicitSpectral2015, vencelsSpectralPlasmaSolverSpectralCode2016, koshkarovMultidimensionalHermitediscontinuousGalerkin2021, filbetConservativeDiscontinuousGalerkin2022},
and have several favorable properties including high accuracy and built-in conservation.
The ion distribution function is approximated with the following representation:
\begin{align}
    f_i(\bm{x}, v_x, v_z) = \sum_{l=0}^{N_{v_x}} \sum_{m=0}^{N_{v_z}} \bm{f}^{lm}_i(\bm{x}) \frac{\exp \left( -\frac{v_x^2 + v_z^2}{2 v_{th}^2} \right) }{2\pi v_{th}^2} He_l \left( \frac{v_x}{v_{th}} \right) He_m \left( \frac{v_z}{v_{th}} \right),
\end{align}
where $He_n$ are the normalized probabilist's Hermite polynomials with weight function $w(\xi) = \frac{1}{\sqrt{2\pi}} e^{-\xi^2 / 2}$.
The semi-discrete system of equations for the Hermite modes $\bm{f}_i^{lm}(\bm{x}, t)$ is
\begin{align}
    \label{eqn:ion-vlasov-semidiscrete}
    \begin{split}
    \partial_t \bm{f}_i^{lm} &+ \left( V^H_{lp} \partial_x \bm{f}^{pm}_i + V^H_{mq} \partial_z \bm{f}^{lq}_i \right) \\
                             &+ \frac{Z_i}{A_i} \frac{\opt}{v_{th}} \left[ E_x D^H_{lp} \bm{f}^{pm}_i + E_z D^H_{mq} \bm{f}^{lq}_i \right] \\
                             &+ \frac{Z_i}{A_i} \oct \left[ -B_0 V_{mq}^H D^H_{lp} \bm{f}^{pq}_i + B_0 V_{lp}^H D^H_{mq} f^{pq}_i \right] = 0,
\end{split}
\end{align}
where we have left sums over repeated indices $p$ and $q$, corresponding to Hermite modes in $v_x$ and $v_z$ respectively, implicit.
The matrices $V^H$ and $D^H$ are tridiagonal matrices whose entries are determined from properties
of the Hermite polynomials.
\cite{coughlinAsymptoticNonasymptoticModel2024}

The spatial discretization of equations \eqref{eqn:ion-vlasov-semidiscrete} and \eqref{eqn:electron-drift-advection} is accomplished with a Fourier pseudospectral discretization in the $x$ direction
and a high-order finite difference scheme in $z$.
Fourier discretizations are highly efficient for periodic domains, achieving spectral accuracy \cite{boydChebyshevFourierSpectral2001}
for smooth solutions. Because the problems solved here do not develop shocks or other discontinuities, the Fourier method is a natural
choice for the periodic $x$ dimension.
On the other hand, a high-order finite difference scheme is a natural choice for the bounded $z$ dimension. 
To illustrate, we write both in the following abstract form:
\begin{align*}
\partial_t \bm{q} + \partial_x (U_x \bm{q}) + \partial_z (U_z \bm{q}) = \bm{S}(\bm{q}),
\end{align*}
where $U_x$ and $U_z$ are linear flux functions (although in the case of the electron drift advection equation they are non-constant in $\bm{x}$).
The $\partial_x$ operator is evaluated using a Fourier pseudospectral collocation scheme,
\begin{align*}
    \partial_x (U_x \bm{q}_h)(x_i, z) = \mathcal{F}^{-1} \left[ \frac{2\pi i k_x}{L_x} \mathcal{F} \{ U_x \bm{q}_h \} (k_x, z) \right] (x_i, z),
\end{align*}
where $\bm{q}_h$ is a grid function evaluated at collocation points $x_i$, $\mathcal{F}$ is the discrete Fourier transform, and $L_x$ is the length of the domain.
Derivatives in $z$ are evaluated using a fifth-order Shu-Osher conservative finite difference method,
\begin{align*}
    \partial_z F(\bm{q}_h) \vert_{z = z_j} = \frac{1}{\Delta z} \left( \hat{F}_{j+1/2} - \hat{F}_{j-1/2} \right).
\end{align*}
The numerical flux $\hat{F}$ is split into left-going and right-going parts which are reconstructed
from upwind-biased stencils:
\begin{align*}
    \hat{F}_{j+1/2} = \hat{F}^+_{j+1/2} + \hat{F}^-_{j+1/2},
\end{align*}
where $\hat{F}^+_{j+1/2} = \mathcal{R}^+(F^+, j+1/2)$ and $\hat{F}^-_{j+1/2} = \mathcal{R}^-(F^-, j+1/2)$
are reconstructed from the splitting of the analytic flux $F = F^+ + F^-$.
Details on the reconstruction stencils can be found in Ref. \onlinecite{vogmanDoryGuestHarris2014}, Equation (20).
We use a purely upwind analytic flux splitting for the electron drift-advection equation and a
Lax-Friedrichs flux splitting for the Hermite $V^H$ operator.
Gauss's law \eqref{eqn:gauss-law} is solved using a sixth-order centered finite
difference stencil in $z$ and a pseudospectral discretization in $x$.

The equations are discretized in time with a third-order four-stage 
Strong-Stability-Preserving \cite{gottliebStrongStabilityPreserving2011}
Runge-Kutta scheme \cite{durranNumericalMethodsFluid2010}.
For an autonomous ordinary differential equation $u'(t) = f(u)$, the scheme is defined as follows:
\begin{align}
    \label{eqn:ssprk43}
    u^1 &= u^n + \frac{\Delta t}{2} f(u^n) \\
    u^2 &= u^1 + \frac{\Delta t}{2} f(u^1) \\
    u^3 &= \frac{2}{3} u^1 + \frac{1}{3} \left[ u^2 + \frac{\Delta t}{2} f(u^2) \right] \\
    u^{n+1} &= u^3 + \frac{\Delta t}{2} f(u^3).
\end{align}

The numerical scheme described here has been benchmarked on continuum kinetic plasma problems in
Ref. \onlinecite{coughlinAsymptoticNonasymptoticModel2024}.

\section{Order-$\epsilon$ Fredholm solvability condition for Landau-Fokker-Planck collision operator}
\label{sec:landau-agyrotropic}

In this section we discuss the satisfiability of \eqref{eqn:f2-kinetic} for the case where the
collision operator $C$ is chosen to be the Landau-Fokker-Planck operator,
\begin{align}
    \label{eqn:landau-fp}
    C(f_s) = \sum_{\sp} C(f_s, f_\sp) = \sum_\sp \nu_\ssp \nabla_{\bm{v}} \cdot \left[ \mathbb{D}_\sp \cdot \nabla_{\bm{v}} f_s - \frac{m_s}{m_\sp} \bm{A}_\sp f_s \right],
\end{align}
where the diffusion tensor $\mathbb{D}_\sp$ and drift vector $\bm{A}_\sp$ can be calculated using the Rosenbluth potentials \cite{rosenbluthFokkerPlanckEquationInverseSquare1957, taitanoMassMomentumEnergy2015}.
The first-order collision term for a bilinear collision operator such as \eqref{eqn:landau-fp}
naturally separates as
\begin{align}
    \label{eqn:landau-fp-split}
    C^1(f_s) = \sum_\sp C(f_s^0, f_\sp^1) + C(f_s^1, f_\sp^0).
\end{align}
Writing the dependence on parallel and perpendicular velocity explicitly, we calculate the first term under the coordinate
transform $\bm{v}_\perp \mapsto -\bm{w}_\perp$:
\begin{align*}
    \left[ C(f_s^0(v_\parallel, \bm{v}_\perp), f_\sp^1(v_\parallel, \bm{v}_\perp) \right](v_\parallel, -\bm{v}_\perp) 
    &= \left[ C(f_s^0(v_\parallel, -\bm{w}_\perp), f_\sp^1(v_\parallel, -\bm{w}_\perp) \right](v_\parallel, \bm{w}_\perp) \\
    &= \left[ C(f_s^0(v_\parallel, \bm{w}_\perp), -f_\sp^1(v_\parallel, \bm{w}_\perp) \right](v_\parallel, \bm{w}_\perp) \\
    &= -\left[ C(f_s^0(v_\parallel, \bm{w}_\perp), f_\sp^1(v_\parallel, \bm{w}_\perp) \right](v_\parallel, \bm{w}_\perp),
\end{align*}
where we have used the fact that $f_\sp^1$ is an odd function of $\bm{v}_\perp$ and bilinearity of $C$.
That is, the first term of \eqref{eqn:landau-fp-split} is an odd function of $\bm{v}_\perp$.
Similarly, the fact that $f_s^1$ is an odd function of $\bm{v}_\perp$ shows that the the second
term of \eqref{eqn:landau-fp-split} is an odd function of $\bm{v}_\perp$.
Thus,
\begin{align*}
    [C^1(f_s)](v_\parallel, -\bm{v}_\perp) = -[C^1(f_s)](v_\parallel, \bm{v}_\perp),
\end{align*}
which shows that $\overline{C^1(f_s) = 0}$.

\end{document}